\documentclass[12pt]{article}
\usepackage{amssymb}
\usepackage{amsfonts}
\usepackage{youngtab}
\usepackage{amsmath,amssymb,latexsym,cite}
\usepackage{amsthm}
\usepackage{cite}
\usepackage[a-1b]{pdfx}    
\usepackage[]{hyperref}
\input xy
\xyoption{all}

\usepackage{hyperref}
\usepackage{pdfsync}
\usepackage{enumitem}
\usepackage[capitalize,nameinlink]{cleveref}
\Crefname{conjecture}{Conjecture}{Conjectures}
\usepackage{youngtab}

\newtheorem{thm}{Theorem}[section]

\newtheorem{remark}[thm]{Remark}

\newcommand{\C}{{\mathbb C}}

\newcommand{\cS}{{\mathcal S}}

\newcommand{\cO}{{\mathcal O}}

\newcommand{\Fl}{\mathrm{Fl}}

\numberwithin{equation}{section}

\begin{document}

\vspace*{0.5in}

\begin{center}

{\large\bf Quantum K theory rings of partial flag manifolds}

\vspace*{0.2in}

Wei Gu$^{1,3}$, Leonardo Mihalcea$^2$, Eric Sharpe$^1$, Weihong Xu$^2$, Hao Zhang$^1$, Hao Zou$^{4,5}$

\begin{tabular}{cc}
{\begin{tabular}{l}
$^1$ Department of Physics, MC 0435\\
850 West Campus Drive\\
Virginia Tech\\
Blacksburg, VA  24061
\end{tabular}}
&
{\begin{tabular}{l}
$^2$ Department of Mathematics, MC 0123 \\
225 Stanger Street\\
Virginia Tech\\
Blacksburg, VA  24061
\end{tabular}}
\end{tabular}

\begin{tabular}{cc}
{\begin{tabular}{l}
$^3$ Max-Planck-Institut f\"ur Mathematik\\
Vivatsgasse 7\\
D-53111 Bonn, Germany
 \end{tabular}}
&
{\begin{tabular}{l}
$^4$  Yau Mathematical Sciences Center\\
Tsinghua University\\
Beijing 100084, China
\end{tabular}}
\end{tabular}

{\begin{tabular}{l}
$^5$  Beijing Institute of Mathematical Sciences and Applications \\
Beijing 101408, China
\end{tabular}}

{\tt weig8@vt.edu}, 
{\tt lmihalce@math.vt.edu},
{\tt ersharpe@vt.edu},
{\tt weihong@vt.edu},
{\tt hzhang96@vt.edu},
{\tt hzou@vt.edu}

$\,$

\end{center}

In this paper we use three-dimensional gauged linear sigma models to
make physical predictions for Whitney-type presentations of
equivariant quantum K theory rings of
partial flag manifolds, as quantum products of universal subbundles
and various ratios, extending previous work for
Grassmannians.  Physically, these arise
as OPEs of Wilson lines for certain Chern-Simons levels.
We also include a simplified method for computing Chern-Simons levels
pertinent to standard quantum K theory.

\begin{flushleft}
June 2023
\end{flushleft}

\newpage

\tableofcontents

\newpage

\section{Introduction}

Recently there has been interest in computations of quantum K theory
rings in mathematics from three-dimensional supersymmetric
gauge theories, see e.g.~\cite[section 2.4]{Bullimore:2014awa},
\cite{Jockers:2018sfl,Jockers:2019wjh,Jockers:2019lwe,
Ueda:2019qhg,Gu:2020zpg,Gu:2022yvj}.  The basic idea is that the quantum
K theory ring of a Fano variety
arises as the OPE ring of Wilson lines in the three-dimensional
gauge theory, which is defined on a three-manifold which has the form
of a circle bundle.  The Wilson lines are the holonomies around the
$S^1$ fibers, and after reducing to two dimensions, the computations of
the Wilson line OPEs reduce to ordinary two-dimensional OPEs,
computable using standard GLSM Coulomb branch methods \cite{Morrison:1994fr}.

In this paper we will apply these ideas to construct Whitney-type
presentations of
equivariant quantum K theory rings of partial flag manifolds,
as quantum products of universal subbundles of the partial flag manifold
and various quotients.  To be clear, we are not the first to study
equivariant quantum K theory of flag manifolds, and in fact,
the equivariant K theory of the total space of the cotangent bundle
of the full (complete) flag manifold arises in integrable systems,
see e.g.~\cite{Bullimore:2014awa,Koroteev:2017nab,Givental:1993nc,kim1,Givental:2001clq},
\cite[section 4.2]{Koroteev:2021lvp}.  
Quantum K theory of the zero section can be obtained
as a limit.  See also \cite{mns,act} for other presentations of the
equivariant quantum K theory rings of partial flag manifolds.
The purpose of this paper is to explore novel presentations -- we do
not claim any novelty in determining the rings themselves.

We begin in section~\ref{sect:genl} 
with a review of the basics of these computations,
which we apply in section~\ref{sect:rev:grass} to the
case of Grassmannians $G(k,n)$.
A presentation of the quantum K theory ring
of $G(k,n)$ in terms of quantum products of bundles (the $\lambda_y$ class
relation) was previously derived on physical grounds in
\cite{Gu:2020zpg}, and later rigorously proven in \cite{Gu:2022yvj}.
We review here a much more efficient version of that analysis,
which will form the prototype of our discussion of flag manifolds.

In section~\ref{sect:flag} we turn to the main content of this paper,
namely proposals for presentations of the equivariant
quantum K theory ring of general partial
flag manifolds.  
We apply Coulomb branch methods to derive a formal expression in terms of
Chern roots of bundles, which we symmetrize into the form of a relation
between $\lambda_y$ classes of universal subbundles and various quotients
thereof.  Our main result is the ring relation~(\ref{eq:qk:flag:reln:lambda0}),
namely
\begin{equation}
\lambda_y( {\cal S}_i) \star
\lambda_y( \widetilde{\mathcal R}_i ) \: = \:
\lambda_y({\cal S}_{i+1}) \: + \:
q_i y^{k_{i+1}-k_i} \left( \det \widetilde{\mathcal R}_i \right) \star
\lambda_y( {\cal S}_{i-1} ),
\end{equation}
where the ${\cal S}_i$ are the universal subbundles on the flag manifold,
and $\widetilde{\mathcal R}_i$ is defined in~(\ref{eq:dictionary}).
This relation can alternatively be expressed without
$\widetilde{\mathcal R}_i$ as~(\ref{eq:qk:flag:reln:lambda}), namely
\begin{equation}\label{E:lambda-y-pres}
\lambda_y({\cal S}_i) \star \lambda_y({\cal S}_{i+1}/{\cal S}_i) \: = \:
\lambda_y({\cal S}_{i+1}) \: - \:
y ^{k_{i+1} - k_{i}} \frac{q_i}{1-q_i} \det ({\cal S}_{i+1}/{\cal S}_i) \star
\left( \lambda_y({\cal S}_{i}) - \lambda_y({\cal S}_{i-1}) \right).
\end{equation}

There are other presentations of the quantum K theory rings of flag manifolds 
in the literature. We already mentioned the presentation from \cite{Gu:2020zpg,Gu:2022yvj}
for Grassmannians, which is a special case of \eqref{E:lambda-y-pres}; in these references we
also found a `Coulomb branch presentation', closely related to one found earlier by Gorbounov
and Korff \cite{gkbethe} in relation to Bethe Ansatz. For the complete flag manifolds, a presentation based on 
integrable systems techniques is stated in \cite{Koroteev:2017nab}; a similar presentation has been recently
proved in \cite{lns,maeno.naito.sagaki:pres1} based on the relation between the 
(equivariant) K theory of complete flag manifolds and the K-theory of semi-infinite flag manifolds. 
Both these presentations are
symmetric in the variables involved, and generalize the usual `Borel presentation' for the ordinary K theory 
of flag manifolds. The presentation given by the relations \eqref{E:lambda-y-pres} has a different
source: it generalizes the presentation arising from the flag manifold seen as a tower of Grassmann
bundles. As such, the relations in \eqref{E:lambda-y-pres} are not symmetric. We refer the reader 
to Section~\ref{sect:ex:f123} for an example of the presentation from \cite{Koroteev:2017nab,maeno.naito.sagaki:pres1} for the quantum ring of 
$\mathrm{Fl}(3)$.

The result above has the correct
classical limit in torus-equivariant K theory, and also correctly specializes
to rigorous mathematics results \cite{Gu:2022yvj} 
for equivariant quantum K theory of Grassmannians.
As another consistency check, in subsection~\ref{sect:cons:qh} we
compare predicted
quantum cohomology ring relations to existing results in the literature.
As another consistency check, in subsection~\ref{sect:fl:dual}
we also demonstrate that this ring relation is consistent with duality
between the flag manifold $F(k_1, \cdots, k_s, N)$ and its
dual $F(N-k_s, \cdots, N-k_1, N)$, and can be most efficiently expressed,
formally, by relating honest bundles to `quantum' bundles, extending 
a similar result for quantum K theory of Grassmannians described
in subsection~\ref{sect:gr:dual}.

Finally, as further consistency tests,
in section~\ref{sect:exs} we compare the predictions for
general partial flag manifolds to existing results in the literature,
namely for incidence varieties (flag manifolds of the form
$F(1,N-1,N)$) and full flag manifolds.
In particular, 
a rigorous proof of our assertions for quantum K theory
for the special case of incidence
varieties will appear in \cite{mathpaper}, as we review in
section~\ref{sect:exs:inc}.

This paper focuses on quantum K theory rings, and not quantum K theory
invariants.  In principle, in small quantum K theory, if we know both the
ring structure (structure constants) as well as a pairing, then we can
compute all of the correlation functions.  In principle, this can be
accomplished in physics using supersymmetric localization.  We will not
pursue this direction here, however.

In passing, we also mention that there exists
(unrelated) work on quantum sheaf cohomology on
partial flag manifolds, see \cite{Guo:2018iyr},
and, separately, there has been work in the physics community on
$I$ functions for vector bundles over Grassmannians and flag manifolds,
relevant for the study of Wilson lines,
see e.g.~\cite{Bonelli:2013mma}.

\section{Review}
\label{sect:genl}

In this section we review the computation of quantum K theory rings from
physical Coulomb branch relations in three-dimensional gauged linear
sigma models.  
One way to do this is to compute for the total space of the cotangent bundle
in a three-dimensional $N=4$ theory, and break to $N=2$; here,
we compute directly in a three-dimensional $N=2$ theory.
We also discuss a (to our knowledge) novel method for
computing Chern-Simons levels relevant to quantum K theory, which replaces
Casimirs of nonabelian groups by computations on Coulomb branches.

Briefly, the idea of 
\cite{Bullimore:2014awa,Jockers:2018sfl,Jockers:2019wjh,Jockers:2019lwe,Gu:2020zpg} 
is that for e.g.~Fano varieties realized in GLSMs,
the quantum K theory ring
can be computed using GLSM Coulomb branches in a manner closely analogous
to 
quantum cohomology rings 
\cite{Morrison:1994fr}.  One starts with a three-dimensional GLSM,
compactifies on a circle of radius $R$, and then considers the compactified
two-dimensional theory obtained by summing the tower of Kaluza-Klein modes.
For GLSMs for Fano spaces,
Coulomb branch computations in the compactified theory yield
quantum K theory relations, just as Coulomb branch computations in an
ordinary two-dimensional GLSM yield quantum cohomology rings,
as in \cite{Morrison:1994fr}.  For such spaces, both the quantum cohomology
ring relations and quantum K theory ring relations can be obtained
in the same way -- from the critical locus of a quantum-corrected
twisted effective superpotential.  The difference is the form of the 
superpotential, which in the original two-dimensional theory involved
ordinary logarithms, and in the compactified theory involve
dilogarithms.

For quantum K theory, the one-loop effective twisted superpotential obtained
by regularizing the infinite sum of KK modes 
has the form
\cite[eq. (2.1)]{Gu:2020zpg}, \cite[equ'n (2.33)]{Closset:2016arn},
\cite[section 2.2.2]{Nekrasov:2009uh}:
\begin{eqnarray} 
    \mathcal{W} & = & \frac{1}{2} k^{ab} (\ln X_a) (\ln X_b)
  \: + \: k^{ai} (\ln X_a) \left(\ln T_i^{-1}\right)
\nonumber \\
    &  & \: + \: \sum_a (\ln q_a) (\ln X_a) \: + \:
 \sum_a \left( i \pi \sum_{\mu {\rm ~pos'}} \alpha_{\mu}^a\right) (\ln X_a)
\nonumber \\
    &  & \: + \: \sum_i \left[
{\rm Li}_2\left(X^{\rho_i}/ T_i\right)  \: + \:
 \frac{1}{4}\left(\ln \left( X^{\rho_i} /T_i \right) \right)^2
 \right],
\label{eq:genl:twisted-sup}
\end{eqnarray}
where Li$_2$ denote the dilogarithm function,
the summation over $i$ is over all the matter fields, 
$T_i$ is the flavor symmetry fugacity and
\begin{equation}
    X^{\rho_i} = \prod_{a} X_a^{Q^i_a}.
\end{equation}
The fields $X_a$ are determined by the Coulomb branch $\sigma$ fields,
in the form
\begin{equation}
X_a \: = \: \exp\left(2 \pi  R \sigma_a\right),
\end{equation}
where $R$ is the radius of the circle on which the three-dimensional
theory was compactified.
The reader should note that the $\sigma_a$, and hence the $X_a$,
are effectively constrained to avoid `excluded loci,' 
and also that even after the gauge group is broken to an abelian subgroup along the Coulomb branch, the Weyl group still acts to interchange the
$\sigma_a$ and $X_a$.  (See 
e.g.~\cite{Hori:2011pd,Gu:2018fpm} for further explanation of these points.)

For the case of gauge group $U(k)$, it can be shown that
\begin{equation}
    i\pi \sum_{\mu~{\rm pos}'} \alpha_\mu^a = i \pi (k-1)
\end{equation}
for all $a$, so the effect will be to multiply $q$ by the phase $(-)^{k-1}$.

It remains to describe the Chern-Simons levels $k^{ab}$ and $k^{ai}$.
Our previous paper \cite{Gu:2020zpg} computed them group-theoretically.
For example, for a nonabelian simple gauge group $G$,
there is a contribution to the pertinent level $k_G$ 
from matter in representation $R$ given by
\begin{equation}
 - \frac{1}{2} T_2 (R),
\end{equation}
where $T_2(R)$ is the quadratic index of the representation $R$, 
normalized so that, in $SU(k)$, $T_2({\bf fund}) = 1$.
(In general,
this may also be combined with a contribution arising from the gauge fields.)
This expression was combined with trace identities in
\cite{Gu:2020zpg} to compute the $k^{ab}$.

In this paper, for computational efficiency, we will use another
expression, obtained in the spirit of the abelian/nonabelian correspondence.
Specifically,
we will compute the levels at generic points along the Coulomb branch,
where the nonabelian gauge symmetry is Higgsed to an abelian
gauge symmetry.  This means in part that
we replace the original gauge group by its Cartan torus,
$U(1)^r$ for $r$ the rank.  
In addition, we must also keep track of the (massive)
W bosons, which we can describe as fields of R charge 2 and $U(1)^r$ charges
given by the root vectors.
Then, to reproduce mathematical results for quantum K theory,
we take the Chern-Simons levels to be
\begin{equation}  \label{eq:genl:cs-level}
k^{ab} \: = \: \frac{1}{2} \sum_{i} (R_i-1) Q^a_{i} Q^b_i,
\end{equation}
where the sum is over all matter fields (including the W bosons),
$R_i$ is the R-charge of the matter field ($0$ for an ordinary field,
$2$ for a W-boson), and $Q^a_i$ is the charge of the field under the
$a$th $U(1)$.  

Briefly, the underlying reason for the equivalence of these two computations
is that, for a simple Lie group,
for $T_R^{\alpha}$ matrices
representing the Lie algebra generators in representation $R$,
\begin{equation}
{\rm Tr} \, T^{\alpha}_R T^{\beta}_R \: = \: \frac{1}{2} T_2(R) \,
 \delta^{\alpha \beta},
\end{equation}
(by definition of quadratic Casimirs, with normalization discussed
earlier)
so if we restrict to the Cartan, and diagonalize the matrices,
then this becomes 
\begin{equation}
\sum_{i=1}^{\dim R} Q_i^a Q_i^b \: = \: \frac{1}{2} T_2(R) \,
 \delta^{ab},
\end{equation}
where the $Q_i^a$ are, simultaneously, the components of the weight
vectors for the representation $R$, and also the $U(1)^r$
charges of the physical
fields in representation $R$, along the Coulomb branch.  
(See also e.g.~\cite[sections 13.2.4, 13.4.2]{DiFrancesco:1997nk}
for related
computations.)

We will apply the same result to flavor symmetries.
We will implicitly check this Chern-Simons level computation against
results obtained from Casimirs as we proceed and compare to previous results;
in all cases, the simpler computation~(\ref{eq:genl:cs-level}) 
will give the same results as
obtained previously from Casimirs.

\section{Warmup: Grassmannians}
\label{sect:rev:grass}

In this section we will describe the computation of the
equivariant quantum K theory ring of a Grassmannian $G(k,n)$,
to prototype the methods we will apply to
partial flag manifolds.
Now, as noted earlier, quantum K theory rings of Grassmannians and
flag manifolds have been computed previously in physics in e.g.~the
integrable systems literature.
Here, we review an improved version of the computations
in \cite{Gu:2020zpg},
as a warmup before developing new presentations of the
quantum K theory rings of partial
flag manifolds.  Specifically, we will compute pertinent
Chern-Simons levels in the spirit of the abelian/nonabelian
correspondence (rather than Casimirs or via three-dimensiona
$N=4$ constructions, as was done previously), 
and also give a direct derivation of the $\lambda_y$ class relations
which were arrived at considerably less directly in \cite{Gu:2020zpg}.

\subsection{Physical realization, Chern-Simons levels and equations of motion}

The GLSM describing a Grassmannian $G(k,n)$ is a
$U(k)$ gauge theory with $n$ fundamentals. 
Proceeding as in section~\ref{sect:genl},
we will compute the equivariant quantum K theory ring 
by computing Coulomb branch relations (equations of motion)
resulting from the twisted one-loop effective superpotential
of the $S^1$-reduced three-dimensional GLSM.

Since the starting point is a three-dimensional theory,
one must pick Chern-Simons levels, and as discussed earlier,
for purposes of computing quantum K theory rings, specific values of the
levels are relevant. 
Those levels were computed previously in \cite{Gu:2020zpg}
using group theory, and can also be understood by starting with a 
three-dimensional $N=4$ theory and breaking to $N=2$,
see e.g.~\cite{Koroteev:2017nab}.
Here, we will instead compute them differently, in the
spirit of the abelian/nonabelian correspondence,
using equation~(\ref{eq:genl:cs-level}),
and then as a consistency check
compare to those obtained in \cite{Gu:2020zpg}.

First, consider the contribution from the fundamentals.
Each fundamental has R-charge $0$, and the $a$th element of a single
fundamental has charge $+1$ under the $a$th $U(1)$, and $0$ under the
others.  Thus, for the fundamentals, 
\begin{equation}
\frac{1}{2} \sum_{i=1}^n \sum_{c = 1}^k (R_{ic}-1) Q^a_{ic} Q^b_{ic}
\: = \:
- \frac{1}{2} \sum_{i=1}^n \sum_{c = 1}^k \delta^a_c \delta^b_c
\: = \:
- \frac{n}{2} \delta^{ab}.
\end{equation}

Next, consider the W-bosons.  The W-boson $W_{\mu \nu}$ (for $\mu \neq \nu$)
has charge
\begin{equation}
Q(W_{\mu \nu})^a \: = \: -\delta^a_{\mu} + \delta^a_{\nu},
\end{equation}
and R-charge $2$, so
\begin{eqnarray}
\lefteqn{
\frac{1}{2} \sum_{\mu \neq \nu} (R_{\mu \nu}-1) Q(W_{\mu \nu})^a
Q(W_{\mu \nu})^b
} \nonumber \\
& = &
 \frac{1}{2} \sum_{\mu \neq \nu} \left( - \delta^a_{\mu} + \delta^a_{\nu} 
\right) \left(
-\delta^b_{\mu} + \delta^b_{\nu} \right),
\\
& = &
 \frac{1}{2} \sum_{\mu \neq \nu} \left(
\delta^a_{\mu} \delta^b_{\mu} + \delta^a_{\nu} \delta^b_{\nu} - 
\delta^a_{\mu} \delta^b_{\nu} - \delta^b_{\mu} \delta^a_{\nu} \right),
\\
& = &
 \frac{1}{2} \left( 2 (k-1) \delta^{ab} \: - 2\left(1 - \delta^{ab} \right)
 \right),
\\
& = &
k \delta^{ab} - 1.
\end{eqnarray}

Putting this together, we find that equation~(\ref{eq:genl:cs-level}) gives
\begin{equation}  \label{eq:gr:k}
k^{ab} \: = \: \frac{1}{2} \sum_{i} (R_i-1) Q^a_{i} Q^b_i
\: = \: - \frac{n}{2} \delta^{ab} \: + \: \left( k \delta^{ab} - 1 \right).
\end{equation}

Now, let us compare to computations in 
our previous paper \cite{Gu:2020zpg}, where the level was computed
using group theory. (See also \cite{Ueda:2019qhg}.)
There, from \cite[equ'n (2.38)]{Gu:2020zpg}, the terms
\begin{equation}
\frac{1}{2} k^{ab} \left( \ln X_a \right) \left( \ln X_b \right)
\end{equation}
are computed as
\begin{equation}
\frac{1}{2} k_{SU(k)} \sum_a \left( \ln X_a \right)^2
\: + \:
\frac{ k_{U(1)} - k_{SU(k)} }{2k} \left(
\sum_a \ln X_a \right)^2,
\end{equation}
or in other words,
\begin{equation}
k^{ab} \: = \: k_{SU(k)} \delta^{ab} \: + \:
\frac{ k_{U(1)} - k_{SU(k)} }{k}.
\end{equation}
From \cite[equ'n (2.36)-(2.37)]{Gu:2020zpg},
\begin{equation}
k_{U(1)} \: = \: - n/2, \: \: \:
k_{SU(k)} \: = \: 
- \frac{n}{2}T_2({\bf\rm fund}) - \frac{1}{2}T_2({\bf\rm adj})
\: = \: k - n/2,
\end{equation}
hence
\begin{eqnarray}
k_{SU(k)} \delta^{ab} \: + \:
\frac{ k_{U(1)} - k_{SU(k)} }{k}
& = &
(k-n/2) \delta^{ab} \: + \:
\frac{ \left[ - n/2 - (k - n/2) \right] }{k},
\\
& = &
(k-n/2) \delta^{ab} \: - \: 1,
\end{eqnarray}
which matches~(\ref{eq:gr:k}), as expected.

To write down the twisted one-loop effective superpotential for the
equivariant quantum K theory of the Grassmannian
$G(k,n)$, we also need 
the gauge-flavor Chern-Simons levels.
These can be computed using the same formula~(\ref{eq:genl:cs-level}),
interpreting the global symmetry group as if it were a gauge symmetry.
Specifically, we now interpret the $n$ fundamentals of $U(k)$ as
a single bifundamental in the $({\bf k}, {\bf \overline{n}})$ of
$U(k) \times U(n)$, of R charge $0$.  
As any W bosons are charged only under one factor,
not both, they do not contribute, and so equation~(\ref{eq:genl:cs-level})
specializes to
\begin{eqnarray}
k^{ai} & = &
\frac{1}{2} \sum_{bj} (R - 1) Q^a_{bj} Q^i_{bj},
\\
& = &
\frac{1}{2} \sum_{bj} (-1) \delta^a_b \delta^i_j,
\\
& = &
- \frac{1}{2}.
\end{eqnarray}

The effective one-loop twisted superpotential~(\ref{eq:genl:twisted-sup})
then becomes
\begin{eqnarray}
\mathcal{W} & = &
\frac{1}{2}\left( k - \frac{n}{2} \right) \sum_a \left( \ln X_a \right)^2
\: - \: \frac{1}{2} \left( \sum_a \ln X_a \right)^2
\nonumber \\
& & 
\: - \:
\frac{1}{2} \left( \sum_{a=1}^k \ln X_a \right) 
\left( \sum_i \ln T_i^{-1} \right)
\: + \:
\left( \ln (-)^{k-1} q \right) \sum_a \ln X_a
\nonumber \\
& & 
\: + \:
\sum_{i=1}^n \sum_{a=1}^k \left[ 
{\rm Li}_2( X_a / T_i ) \: + \: \frac{1}{4} \left( \ln(X_a / T_i) \right)^2
\right],
\\
& = &
\frac{k}{2}  \sum_a \left( \ln X_a \right)^2
\: - \: \frac{1}{2} \left( \sum_a \ln X_a \right)^2
\nonumber \\
& & 
\: + \:
\left( \ln (-)^{k-1} q \right) \sum_a \ln X_a
\: + \:
\sum_{i=1}^n \sum_{a=1}^k {\rm Li}_2( X_a / T_i )
\: + \: \frac{k}{4} \sum_i \left( \ln T_i \right)^2.
\end{eqnarray}

The equations of motion (Coulomb branch equations) can be
calculated from
\begin{equation} \label{eq:coulomb}
    \exp\left(\frac{\partial \mathcal{W}}{\partial \ln X_{a}}\right) = 1.
\end{equation}
The resulting equations of motion are
\begin{equation}  \label{eq:gr:eom}
(-)^{k-1} q X_a^{k} 
\: = \:
\left( \prod_b X_b \right)
\left( \prod_i ( 1 - X_a / T_i ) \right),
\end{equation}
which matches \cite[equ'n (35)]{Gu:2022yvj},
and is the equivariant extension of \cite[equ'n (2.40)]{Gu:2020zpg}.
These are also the same as the Bethe ansatz equations of 
\cite[equ'n (4.17)]{gkbethe}.
The reader should bear in mind that the $X_a$ are constrained to be
distinct (as coincident $X_a$ lie within the `excluded locus'),
and that there is a residual $S_k$ of the original gauge group $U(k)$
acting that interchanges the $X_a$, as explained in
e.g.~\cite{Hori:2011pd,Gu:2018fpm}.  We will not belabor these points
in this analysis.

\subsection{Characteristic polynomial and symmetrization}
\label{sect:gr:symm}

Now, for purposes of comparing to mathematical results,
the equations of motion~(\ref{eq:gr:eom}) need to be reworked
into a more symmetric form
in order to compare to mathematics results.  
A proposal for a presentation in terms of $\lambda_y$ classes
was given in \cite{Gu:2020zpg,Gu:2022yvj}, through a combination
of physics arguments in shifted variables and rigorous mathematics proofs
using unrelated methods.  We shall describe next how
to obtain that presentation directly, in a fashion that will
generalize to flag manifolds.

To that end, we first rewrite the equations of motion~(\ref{eq:gr:eom})
in the form
\begin{eqnarray}
\lefteqn{
(-)^{k-1} q \left( X_a \right)^k e_n(T) 
} \nonumber \\
& = &
(-)^n e_k(X) \prod_i \left( X_a - T_i \right),
\\
& = & (-)^n e_k(X) \left[ \left( X_a \right)^n \: - \:
e_1(T) \left( X_a\right)^{n-1} \: + \: e_2(T) \left( X_a\right)^{n-2}
\: + \: \cdots \: + \:
(-)^n e_n(T) \right],
\nonumber
\end{eqnarray}
where $e_{\ell}(x)$ denotes the $\ell$th elementary symmetric polynomial
in indeterminates $\{ x_a \}$.
We can rearrange this to the form
\begin{equation}  \label{eq:gr:char}
\sum_{\ell=0}^n (-)^{\ell} \xi^{n-\ell} \left[
e_k(X) \, e_{\ell}(T) \: + \: q \, e_n(T) \, e_{\ell -n+k}(0) \right] \: = \: 0,
\end{equation}
for $\xi = X_a$,
in the convention that $e_{\ell}(0) = \delta_{\ell,0}$.
In the previous work \cite{Gu:2020zpg}, this was referred to as the
characteristic polynomial.

\begin{remark} The Coulomb branch equations \eqref{eq:coulomb} from this note 
and those in \cite{Gu:2020zpg,Gu:2022yvj} coincide. However, the symmetrization 
procedure in this paper differs from the one in {\em loc.cit.}, and it gives a 
{\em different} characteristic polynomial. For example, 
the non-equivariant polynomial of $G(2,4)$ from \cite{Gu:2020zpg,Gu:2022yvj} is
\[ \begin{split} \xi^4+(X_1 X_2-X_1-X_2-3)\xi^3+(-3X_1 X_2+3X_1+3X_2+3)\xi^2+\\
(3X_1X_2
+q-3X_1-3X_2-1)\xi-X_1X_2+X_1+X_2  \/, \end{split} \]
while the polynomial from \eqref{eq:gr:char} above is (after making $T_i =1, 1 \le i \le 4$): 
\[ \xi^4 X_1X_2-4\xi^3 X_1 X_2+\xi^2(6X_1X_2+q)-4\xi X_1X_2+X_1X_2 \/. \]
\end{remark}
Since the characteristic polynomial is $n$th order, it has $n$ roots,
which include the $k$ values of $X_a$, as well as 
$n-k$ additional roots we shall label $\overline{X}_a$.
We let $w$ denote the combined collection $\{ X_a, \overline{X}_{a'} \}$.

Now, we can simplify the characteristic polynomial using
Vieta's formula, which says that for any order $n$ polynomial $P(x)$
\begin{equation}
P(x) \: = \: a_n x^n \: + \: a_{n-1} x^{n-1} \: + \: \cdots \: + \: a_1 x
\: + \: a_0,
\end{equation}
with roots $r_1, \cdots, r_n$,
the coefficients $a_{\ell}$ are related to the roots by
\begin{equation}  \label{eq:vieta}
e_{\ell}(r) \: = \: (-)^{\ell} \, \frac{a_{n-\ell}}{a_n},
\end{equation}
where $r = \{ r_i \}$ denotes the collection of roots.
In the case of the characteristic polynomial~(\ref{eq:gr:char}),
\begin{equation}
a_{n-\ell} \: = \: (-)^{\ell} \left[ e_k(X) \, e_{\ell}(T)
\: + \: q \, e_n(T) \, e_{\ell-n+k}(0) \right],
\end{equation}
so Vieta's formula implies
\begin{equation}  \label{eq:gr:Coulomb:1}
e_k(X) e_{\ell}(w) \: = \: e_k(X) \, e_{\ell}(T) \: + \: q \, e_n(T) \, e_{\ell-n+k}(0).
\end{equation}

Now, let us simplify~(\ref{eq:gr:Coulomb:1}).
First, if $\ell \neq n-k$, then~(\ref{eq:gr:Coulomb:1}) immediately implies
\begin{equation}  \label{eq:gr:w=t}
e_{\ell}(w) \: = \: e_{\ell}(T).
\end{equation}

For $\ell = n-k$, equation~(\ref{eq:gr:Coulomb:1}) implies
\begin{equation}  \label{eq:gr:int2}
e_k(X) \, e_{n-k}(w) \: = \: e_k(X) \, e_{n-k}(T) \: + \: q \, e_n(T).
\end{equation}
We also know that
\begin{equation}  \label{eq:gr:expand-w}
e_{\ell}(w) \: = \: \sum_{r=0}^{n-k} e_{\ell-r}(X) \, e_r(\overline{X}),
\end{equation}
for $\ell = 0, \cdots, n$.
(This follows from general properties of symmetric polynomials, plus
the fact that $e_r(\overline{X}) = 0$ for $r > n-k$, as there are only
$n-k$ $\overline{X}$ indeterminates.)
In particular,
\begin{equation}
e_n(w) \: = \: \sum_{r=0}^{n-k} e_{n-r}(X) \, e_r(\overline{X}),
\end{equation}
but $e_{n-r}(X) = 0$ for $n-r > k$ (meaning, $r < n-k$) as there are
only $k$ $X$ indeterminates.  Thus, using the above and also~(\ref{eq:gr:w=t}),
we can write
\begin{equation} \label{eq:gr:nt=nw}
e_n(T) \: = \: e_n(w) \: = \: e_k(X) \, e_{n-k}(\overline{X}).
\end{equation}
Thus, equation~(\ref{eq:gr:int2}) implies
\begin{equation} \label{eq:gr:interp1}
e_{n-k}(w) \: = \: e_{n-k}(T) \: + \: q \, e_{n-k}(\overline{X}).
\end{equation}
Putting these together, we have
\begin{equation}
e_{\ell}(w) \: = \: \left\{ \begin{array}{cl}
e_{\ell}(T) & \ell \neq n-k,
\\
e_{n-k}(T) +  q \, e_{n-k}(\overline{X}) & \ell = n-k.
\end{array} \right.
\end{equation}
For later use, from~(\ref{eq:gr:interp1}) we have
\begin{equation}
\sum_{r=0}^{n-k} e_{n-k-r}(X) \, e_r(\overline{X}) 
\: = \: e_{n-k}(w) \: = \: e_{n-k}(T) \: + \: q \, e_{n-k}(\overline{X}),
\end{equation}
hence
\begin{equation}
\sum_{r=0}^{n-k-1} e_{n-k-r}(X) \, e_r(\overline{X}) \: + \: e_{n-k}(\overline{X})
\: = \: e_{n-k}(T) \: + \: q \, e_{n-k}(\overline{X}),
\end{equation}
which can be rearrangd to
\begin{equation} \label{eq:gr:interp2}
\sum_{r=0}^{n-k-1} e_{n-k-r}(X) \, e_r(\overline{X}) \: + \:
(1-q) e_{n-k}(\overline{X}) \: = \: e_{n-k}(T).
\end{equation}

Returning to~(\ref{eq:gr:Coulomb:1}),
applying~(\ref{eq:gr:nt=nw}) and cancelling out a common factor of
$e_k(X)$, we have
\begin{equation}
e_{\ell}(w) \: = \: e_{\ell}(T) \: + \: q \, e_{n-k}(\overline{X})
e_{\ell-n+k}(0).
\end{equation}
Applying~(\ref{eq:gr:expand-w}) we then have 
\begin{equation}  \label{eq:gr:Coulomb:final}
\sum_{r=0}^{n-k} e_{\ell-r}(X) e_r(\overline{X})
 \: = \:
e_{\ell}(T) \: + \: q \, e_{n-k}(\overline{X}) \, e_{\ell-n+k}(0).
\end{equation}

Equation~(\ref{eq:gr:Coulomb:final}) is the key result from Vieta's equation.
Next, we solve it algebraically.

To that end, note that equation~(\ref{eq:gr:Coulomb:final}) is
the degree-$\ell$ piece of
\begin{equation}
\label{eq:gr:generating}
\left( \sum_{r=0}^k y^r e_r(X) \right)
\left( \sum_{t=0}^{n-k} y^t e_t( \overline{X} ) \right)
\: = \:
\sum_{r=0}^{n} y^r e_r(T)
\: + \: q y^{n-k} e_{n-k}(\overline{X}),
\end{equation}
which implies
\begin{equation}
 \sum_{r=0}^{n-k} y^r e_r( \overline{X} )
\: = \:
\left( \sum_{r=0}^{n} y^r e_r(T) \right) \left( 
\sum_{t=0}^{\infty} (-)^t y^t h_t(X) \right)
\: + \:
q y^{n-k} e_{n-k}(\overline{X})
\left( 
\sum_{t=0}^{\infty} (-)^t y^t h_t(X) \right),
\end{equation}
for $h_t(X)$ the complete homogeneous symmetric polynomial of degree $t$ in 
$\{X\}$.  
We read off that for $\ell < n-k$,
\begin{equation}
e_{\ell}(\overline{X}) \: = \: \sum_{r=0}^{n-k} (-)^r e_{\ell-r}(T) \, h_r(X),
\end{equation}
and for $\ell = n-k$,
\begin{equation}
e_{n-k}(\overline{X}) \: = \: \sum_{r=0}^{n-k} (-)^r e_{n-k-r}(T) \, h_r(X)
\: + \: q e_{n-k}(\overline{X}),
\end{equation}
hence
\begin{equation}
e_{n-k}(\overline{X}) \: = \: 
(1-q)^{-1} \sum_{r=0}^{n-k} (-)^r e_{n-k-r}(T) \, h_r(X).
\end{equation}

To simplify the expression above, define a $(n-k)$-element
collection $\{ \hat{X} \}$ by
\begin{equation}  \label{eq:gr:defn:hatx}
e_{\ell}(\hat{X}) \: = \:
\sum_{r=0}^{n-k} (-)^r e_{\ell-r}(T) \, h_r(X),
\end{equation}
then we can write
\begin{equation}
e_{\ell}(\overline{X}) \: = \: \left\{ \begin{array}{cl}
e_{\ell}(\hat{X}) & \ell < n-k,
\\
(1-q)^{-1} e_{n-k}(\hat{X}) & \ell = n-k.
\end{array} \right.
\end{equation}

\subsection{Interpretation in terms of bundles:
$\lambda_y$ class presentation}
\label{sect:gr:lambda-pres}

In this section we will interpret~(\ref{eq:gr:Coulomb:final}) in terms of
$\lambda_y$ classes.  (We emphasize at the start that any interpretation
may be slightly ambiguous; we will utilize comparisons to mathematics to
justify our proposal.)

First, for reasons described earlier and elsewhere,
we associate the $\{ X \}$ with Chern roots of the universal
subbundle ${\cal S}$, meaning that
\begin{equation}
e_{\ell}(X) \: \sim \: \wedge^{\ell} {\cal S}.
\end{equation}
In addition, it is natural to associate
the $\{ \hat{X} \}$ with Chern roots of the universal quotient bundle
${\mathbb C}^n/{\cal S}$, meaning
\begin{equation}
e_{\ell}(\hat{X}) \: \sim \: \wedge^{\ell} ( {\mathbb C}^n / {\cal S} ).
\end{equation}
Classically, this follows from
the defining property~(\ref{eq:gr:defn:hatx}) of the $\{ \hat{X} \}$.
In more detail, this follows from the short exact sequence
\begin{equation}
0 \: \longrightarrow \: {\cal S} \: \longrightarrow \: 
{\mathbb C}^n \: \longrightarrow \: {\mathbb C}^n/{\cal S} \:
\longrightarrow \: 0,
\end{equation}
which implies
\begin{equation}
c({\cal S}) \, c( {\mathbb C}^n/{\cal S}) \: = \: c( {\mathbb C}^n),
\end{equation}
and in K theory,
\begin{equation}
\lambda_y( {\cal S}) \, \lambda_y( {\mathbb C}^n/{\cal S}) \: = \: 
\lambda_y( {\mathbb C}^n)
\end{equation}
(where 
$\lambda_y({\cal E}) = 1 + y {\cal E} + y^2 \wedge^2 {\cal E} + \cdots$),
and hence the relation~(\ref{eq:gr:defn:hatx}), after algebra.
(This association will be justified by the fact that this
will correctly reproduce rigorous results for quantum K theory
ring presentations \cite{Gu:2022yvj} and also quantum cohomology rings.)

We formally associate the $\{ \overline{X} \}$ with Chern roots of a bundle
$\widetilde{\mathcal Q}$ of rank $n-k$.
From the results of the last section, we can identify
\begin{equation}  \label{eq:gr:tildeq}
e_{\ell}(\overline{X}) \: \sim \: \wedge^{\ell} \widetilde{\mathcal Q}
\: = \: \left\{ \begin{array}{cl}
\wedge^{\ell} ({\mathbb C}^n/{\cal S}) & \ell < n-k,
\\
(1-q)^{-1} \wedge^{\ell} ( {\mathbb C}^n/{\cal S} ) & \ell = n-k.
\end{array} \right.
\end{equation}
The reader should note that this means that for $q \neq 0$,
$\widetilde{\mathcal Q}$ is not a classical bundle, but instead appears to be
more nearly some sort of quantum exterior product which we shall not
attempt to define carefully here.

With this dictionary in mind, we can now interpret~(\ref{eq:gr:Coulomb:final}),
namely
\begin{equation}  
\sum_{r=0}^{n-k} e_{\ell-r}(X) e_r(\overline{X})
 \: = \:
e_{\ell}(T) \: + \:  q \, e_{n-k}(\overline{X}) \, e_{\ell-n+k}(0),
\end{equation}
as
\begin{equation}
\sum_{r=0}^{n-k} \wedge^{\ell-r}({\cal S}) \star
\wedge^{r} \widetilde{\mathcal Q} \: = \:
\wedge^{\ell}( {\mathbb C}^n ) \: + \:
q \delta_{\ell,n-k} \det \widetilde{\mathcal Q},
\end{equation}
or more elegantly,
\begin{equation}  \label{eq:gr:quantumre}
\lambda_y({\cal S}) \star \lambda_y( \widetilde{\mathcal Q}) \: = \:
\lambda_y( {\mathbb C}^n ) \: + \: y^{n-k} q \det \widetilde{\mathcal Q}.
\end{equation}
Equation~(\ref{eq:gr:quantumre}) is the $\lambda_y$ class relation
defining the $T$-equivariant quantum K theory ring of the
Grassmannian $G(k,n)$.

We can give an alternate presentation that does not involve
$\widetilde{\mathcal Q}$ as follows.
Returning to~(\ref{eq:gr:Coulomb:final}) and again using the
dictionary~(\ref{eq:gr:tildeq}), we have
\begin{eqnarray}
\lefteqn{
\sum_{r=0}^{n-k-1} \wedge^{\ell-r}({\cal S}) \, \star \,
\wedge^r ({\mathbb C}^n/{\cal S}) \: + \:
\frac{1}{1-q}
\wedge^{\ell-(n-k)} {\cal S} \, \star \, \det ({\mathbb C}^n/{\cal S})
} \nonumber \\
& \hspace*{1in} = &
\wedge^{\ell} {\mathbb C}^n \: + \: \frac{1}{1-q} \det( {\mathbb C}^n/{\cal S})
\delta_{\ell, n-k},
\end{eqnarray}
which can be rearranged to the form
\begin{equation}
\sum_{r=0}^{n-k} \wedge^{\ell-r} {\cal S} \, \star \, 
\wedge^r( {\mathbb C}^n/{\cal S}) 
\: = \:
\wedge^{\ell} {\mathbb C}^n \: - \:
\frac{q}{1-q} \det( {\mathbb C}^n/{\cal S}) \, \star \,
\left( \wedge^{\ell - n + k} {\cal S} \: - \:
{\cal O} \delta_{\ell, n-k} \right).
\end{equation}
Adding factors of $y$, this becomes the $\lambda_y$ relation
\begin{equation}  \label{eq:gr:lambda}
\lambda_y({\cal S}) \star \lambda_y( {\mathbb C}^n/{\cal S})
\: = \:
\lambda_y( {\mathbb C}^n) \: - \:
y^{n-k} \frac{q}{1-q} \det({\mathbb C}^n/{\cal S}) \star
\left( \lambda_y({\cal S}) - 1 \right).
\end{equation}

Equation~(\ref{eq:gr:lambda}) is a second form of
the $\lambda_y$ class relation
defining the $T$-equivariant quantum K theory ring of the
Grassmannian $G(k,n)$.  It was argued, less efficiently, in
\cite{Gu:2020zpg}, and proven rigorously using different methods
in \cite{Gu:2022yvj}.  Our analysis of equivariant quantum K theory rings
of flag manifolds, though more complex, will be closely analogous to
that we have demonstrated here for Grassmannians.

\subsection{Duality: $G(k,n)\cong G(n-k,n)$}
\label{sect:gr:dual}

In this section we discuss how our presentation of the quantum K theory
ring of the Grassmannian behaves under the duality of
Grassmannians that relates $G(k,n)$ to the dual $G(n-k,n)$.
This
relates the universal subbundle ${\cal S}$, quotient bundle
${\mathcal Q} = {\mathbb C}^n/{\cal S}$ and vector space ${\mathbb C}^n$ of the
Grassmannian $G(k,n)$ to the ${\cal S}'$, ${\mathcal Q}'$ and $({\mathbb C}^n)'$ of
the dual Grassmannian $G(n-k,n)$ classically as
\begin{equation}
{\cal S}' \: = \: {\mathcal Q}^*, \: \: \:
{\mathcal Q}' \: = \: {\cal S}^*, \: \: \:
({\mathbb C}^n)' \: = \: ( {\mathbb C}^n)^*.
\end{equation}

First, we recall that this mathematical duality is realized
physically as an IR duality of the gauge theories:
$G(k,n)$ is realized by a $U(k)$ gauge theory with $n$ fundamentals and 
Chern-Simons levels 
\begin{equation}
\label{eq:gr:cslevels}
    k^{ab} = -\frac{n}{2}\delta^{ab} + (k\delta^{ab}-1) \quad \text{and}\quad k^{ai} = -\frac{1}{2}\, ,
\end{equation}
while the dual Grassmannian $G(n-k,n)$ is realized by
a $U(n-k)$ gauge theory with $n$ fundamentals and 
Chern-Simons levels 
\begin{equation}
    k^{ab} = -\frac{n}{2}\delta^{ab} + ((n-k)\delta^{ab}-1) \quad \text{and}\quad k^{ai} = -\frac{1}{2}\, 
\end{equation}
(replacing $k$ by $n-k$ in Equation~(\ref{eq:gr:cslevels})). 

Now, we turn to the mathematics.
If we let quantities on the dual Grassmannian be denoted with a prime
($\prime$), then the $\lambda_y$ relation~(\ref{eq:gr:lambda})
on the dual Grassmannian is
\begin{equation}
\lambda_{y'}({\cal S}') \star
\lambda_{y'}( ({\mathbb C}^n)^*/{\cal S}')
\: = \:
\lambda_{y'}( ({\mathbb C}^n)^*) \: - \:
(y')^k \frac{q'}{1-q'} \det( ({\mathbb C}^n)^*/{\cal S}') \star
\left( \lambda_{y'}({\cal S}') - 1 \right).
\end{equation}
Reinterpreting the quantities above on the original Grassmannian,
this is the relation
\begin{equation}
\lambda_{y'}({\mathcal Q}^*) \star \lambda_{y'}({\cal S}^*)
\: = \:
\lambda_{y'}( ({\mathbb C}^n)^* ) \: - \:
(y')^k \frac{q'}{1-q'} \det({\cal S}^*) \star
\left( \lambda_{y'}({\mathcal Q}^*) - 1 \right).
\end{equation}
As a consistency check, note the classical limit is correct,
by virtue of the dual of the
sequence defining the universal subbundle ${\cal S}$.
We will not use this relation in this paper, but include it for completeness.

\section{Partial flag manifolds}
\label{sect:flag}

In this section we compute the equivariant quantum K theory rings
of partial flag manifolds.  Although the details are considerably
more complicated, the underlying logic is the same as that
presented in section~\ref{sect:rev:grass} for Grassmannians.
We should add that our focus is on deriving novel presentations;
the underlying Coulomb branch description of the quantum K theory
ring has been previously obtained from the corresponding Bethe ansatz,
for example, see e.g.~\cite{Koroteev:2017nab}, and other presentations
of the quantum K theory ring have been described in mathematics in
e.g.~\cite{Givental:2001clq,mns,act}.

\subsection{Physical realization and twisted effective superpotential}
\label{sect:flag:physics}

The flag manifold $F(k_1, \ldots, k_s;n)$ is realized physically as
a $U(k_1) \times U(k_2) \times \ldots \times U(k_{s})$ gauge theory with matter fields which are bifundamentals in the $(\mathbf{k_i}, \mathbf{\overline{k_{i+1}}})$ representation of $U(k_i) \times U(k_{i+1})$ for $i = 1, \ldots, s-1$ and $n$ fundamentals of $U(k_s)$, see e.g.~\cite{Donagi:2007hi}. 
We will use $\Phi^{a_i,a_{i+1}}$ 
$(a_i = 1, \ldots, k_i; a_{i+1} = 1, \ldots, k_{i+1})$ 
to denote the bifundamental of $U(k_i) \times U(k_{i+1})$. 
We will also use the notation $\Phi^{a_s, j}$ ($j= 1, \ldots, n$) 
to denote the fundamental of $U(k_s)$. We will use the convention $k_0 = 0$.

Generically along the Coulomb branch,
each $U(k_i)$ is broken to $U(1)^{k_{i}}$, 
giving rise to $k_i(k_i-1)$ W-bosons 
$W^{m_i,n_i}$ ($m_i, n_i = 1, \ldots, k_i$). 
The fields charged under the 
$b_i$-th $U(1)$ factor of $U(1)^{k_i}$
are
$\Phi^{a_i, a_{i+1}}$, $\Phi^{a_{i-1}, a_i}$ and $W^{m_i,n_i}$,
with charges as listed below: 
\begin{center}
\begin{tabular}{c|c}
Field & Charge \\ \hline
$\Phi^{a_i, a_{i+1}}$ & $Q^{a_i, a_{i+1}}_{b_i} = +\delta^{a_i}_{b_i}$ \\
$\Phi^{a_{i-1}, a_i}$ & $Q_{b_i}^{a_{i-1},a_i} = - \delta^{a_i}_{b_i}$ \\
$W^{m_i, n_i}$ & $Q^{m_i, n_i}_{b_i} = - \delta^{b_i}_{m_i} + \delta^{b_i}_{n_i}$
\end{tabular}
\end{center}

Also, associated to 
the $b_i$-th $U(1)$ factor is a sigma field we denote $\sigma^{(i)}_{b_i}$,
encoded in $X^{(i)}_{b_i}$.

Then, the effective twisted superpotential~(\ref{eq:genl:twisted-sup})
is given by
\begin{equation} \label{eq:w-flag-0}
\mathcal{W} \: = \: \sum_{i=1}^s \mathcal{W}_i,
\end{equation}
where
\begin{eqnarray}
\mathcal{W}_i & = &
\frac{1}{2} \sum_{j=1}^s \sum_{a_i = 1}^{k_i} \sum_{b_j =1}^{k_j}
k^{a_i b_j} \left( \ln X^{(i)}_{a_i} \right) \left( \ln X^{(j)}_{b_j} \right)
\: + \:
\left( \ln \left( (-)^{k_i-1} q_i \right) \right)
\sum_{a_i=1}^{k_i} \ln X^{(i)}_{a_i} 
\nonumber \\
& & 
\: + \:
\sum_{a_i=1}^{k_i} \sum_{a_{s+1}=1}^n k^{a_i a_{s+1}} 
\left( \ln X^{(i)}_{a_i} \right) \left( \ln T_{a_{s+1}}^{-1} \right)
\nonumber \\
& & 
\: + \:
\sum_{a_i=1}^{k_i} \sum_{a_{i+1}=1}^{k_{i+1}} \left[
{\rm Li}_2\left( X^{(i)}_{a_i} / X^{(i+1)}_{a_{i+1}} \right)
\: + \:
\frac{1}{4} \left( \ln \left( X^{(i)}_{a_i} / X^{(i+1)}_{a_{i+1}} \right)
\right)^2 \right],
\end{eqnarray}
in conventions where $k_0 = 0$, $k_{s+1} = n$,
and $X^{(s+1)}_{a_{s+1}} = T_{a_{s+1}}$.

It remains to compute the Chern-Simons levels,
using equation~(\ref{eq:genl:cs-level}).

First, for the first gauge factor, we have
\begin{eqnarray}
k^{a_1 b_j} & = &
- \frac{1}{2} \sum_{c_1, c_{2}} Q^{c_1,c_{2}}_{a_1} Q^{c_1,c_2}_{b_j}
\: + \: \frac{1}{2} \sum_{m_1 \neq n_1 = 1}^{k_1} Q^{m_1,n_1}_{a_1} Q^{m_1,n_1}_{b_j},
\\
& = & \delta_{1,j} \left(
- \frac{k_2}{2} \delta^{a_1 b_j} \: + \:  k_1 \delta^{a_1 b_j} - 1\right)
\: + \: \frac{1}{2}\delta_{2,j}.
\end{eqnarray}

Then, for $1 < i,j < s$, the only nonzero levels $k^{a_i b_j}$ have
$j \in \{i-1,i,i+1\}$.  For those values,
\begin{equation}
\begin{aligned}
    k^{a_i b_j} &
= -\frac{1}{2} \sum_{c_i} \sum_{c_{i+1}} Q^{c_i, c_{i+1}}_{a_i} Q^{c_j, c_{j+1}}_{b_j} 
- \frac{1}{2}\sum_{c_{i-1}}\sum_{c_i} Q^{c_{i-1}, c_i}_{a_i} Q^{c_{j-1},c_j}_{b_j} 
+ \frac{1}{2} \sum_{m_i \neq n_i = 1}^{k_i}  Q^{m_i, n_i}_{a_i} Q^{m_i, n_i}_{b_j}, 
\\
    &= \delta_{i,j} \left( - \frac{k_{i+1}}{2}  \delta^{a_i b_i} - \frac{k_{i-1}}{2} \delta^{a_i b_i} + 
    k_i\delta^{a_i b_i} - 1 \right)
\: + \: \frac{1}{2}\left( \delta_{j,i+1} \: + \:  \delta_{j,i-1} \right),
\\
    &= \delta_{i,j} \left[ \left(k_i - \frac{k_{i-1}}{2} - \frac{k_{i+1}}{2}\right) \delta^{a_i b_i} - 1 \right]
\: + \:
 \frac{1}{4} \left( \delta_{j,i+1} \: + \:  \delta_{j,i-1}
\: + \: \delta_{i,j+1} \: + \: \delta_{i,j-1} \right),
\end{aligned}
\end{equation}
where in the last line we have written the last two terms in an
explicitly symmetric fashion.
Note that the levels $k^{a_1 b_i}$ are a special case if we define
$k_{-1} = 0$.

For the last gauge factor, we have
\begin{eqnarray}
k^{a_s b_i} & = &
- \frac{1}{2} \sum_{c_{s-1}, c_{s}} Q^{c_{s-1},c_{2}}_{a_s} Q^{c_{s-1},c_s}_{b_i}
\: + \: \frac{1}{2} \sum_{m_s \neq n_i = 1}^{k_1} Q^{m_s,n_i}_{a_s} Q^{m_s,n_i}_{b_i},
\\
& = & \delta_{s,i} \left(
- \frac{k_{s-1}}{2} \delta^{a_s b_i} \: + \: k_s \delta^{a_s b_i} - 1
\right).
\end{eqnarray}
Finally, the gauge-flavor Chern-Simons level can be computed in
the same fashion as in section~\ref{sect:rev:grass}:
\begin{equation}
k^{a_i j} \: = \: - \frac{1}{2} \delta_{i,s}.
\end{equation}

Next, we plug these levels into the expression~(\ref{eq:w-flag-0})
and simplify to find that
the full superpotential is 
\begin{eqnarray}
\mathcal{W} & = &
\frac{1}{2} \sum_{i=1}^s (k_i - 1)  \sum_{a_i = 1}^{k_i} 
\left(\ln X^{(i)}_{a_i}\right)^2
\: - \:
\sum_{i=1}^s  \sum_{1 \le a_i < b_i \le k_i} \left(\ln X^{(i)}_{a_i}\right) 
\left(\ln X^{(i)}_{b_i}\right)
\nonumber \\
& &
\: + \:
\sum_{i=1}^s  \left(\ln \left( (-)^{k_i-1}q_i\right)\right) 
\sum_{a_i = 1}^{k_i} \left(\ln X^{(i)}_{a_i}\right)
\nonumber \\
& & 
\: + \: \sum_{i=1}^s
\sum_{a_i = 1}^{k_i} \sum_{a_{i+1} = 1}^{k_{i+1}} 
{\rm Li}_2\left(X^{(i)}_{a_i} / X^{(i+1)}_{a_{i+1}}\right) 
\label{eq:W:full}
\end{eqnarray}
in the conventions that $k_{0} = 0$, $k_{s+1} = n$, and
$X^{(s+1)}_{a_{s+1}} = T_{a_{s+1}}$.

\subsection{Coulomb branch equations}

As discussed earlier, the Coulomb branch equations can be calculated from
derivatives of the superpotential~(\ref{eq:coulomb}), which we repeat
below:
\begin{equation} 
    \exp\left(\frac{\partial \mathcal{W}}{\partial \ln X^{(i)}_{a_i}}\right) = 1.
\end{equation}
(They can also be obtained as a limit of e.g.~Bethe ansatz computations,
see e.g.~\cite{Koroteev:2017nab}; our ultimate goal is proposals for
new ring presentations, in this section we are merely reviewing
routes through the physics.)
Computationally, to evaluate the expression above, it is helpful
to collect all of the terms involving just the $X^{(i)}$ (associated
with the $i$th gauge group factor), which are given  
below:
\begin{eqnarray} \label{eq:wi:final}
\lefteqn{
 \frac{1}{2} (k_i - 1)  \sum_{a_i = 1}^{k_i} 
\left(\ln X^{(i)}_{a_i}\right)^2
 - \sum_{1 \le a_i < b_i \le k_i} \left(\ln X^{(i)}_{a_i}\right) 
\left(\ln X^{(i)}_{b_i}\right) 
+ \left(\ln \left( (-)^{k_i-1}q_i\right)\right) 
\sum_{a_i = 1}^{k_i} \left(\ln X^{(i)}_{a_i}\right)
} \nonumber \\
  & \hspace*{0.75in} &
+ \sum_{a_i = 1}^{k_i} \sum_{a_{i+1} = 1}^{k_{i+1}} 
{\rm Li}_2\left(X^{(i)}_{a_i} / X^{(i+1)}_{a_{i+1}}\right) 
+ \sum_{a_{i-1} = 1}^{k_{i-1}} \sum_{a_i = 1}^{k_i}
 {\rm Li}_2\left(X^{(i-1)}_{a_{i-1}} / X^{(i)}_{a_i}\right) 
\end{eqnarray}
in the conventions that $k_{0} = 0$, $k_{s+1} = n$, and
$X^{(s+1)}_{a_{s+1}} = T_{a_{s+1}}$.

Plugging into~(\ref{eq:coulomb}), we have
\begin{equation}\label{eqn:QK-rel}
    (-)^{k_i - 1} q_i \left(X^{(i)}_{a_i}\right)^{k_i} 
\prod_{b_{i-1}=1}^{k_{i-1}} \left(1 - \frac{X^{(i-1)}_{b_{i-1}}}{X^{(i)}_{a_i}}\right)
\: =  \:
\left(\prod_{b_i = 1}^{k_i} X^{(i)}_{b_i} \right)
\prod_{b_{i+1} = 1}^{k_{i+1}} \left(1 - 
\frac{X^{(i)}_{a_i}}{X^{(i+1)}_{b_{i+1}}}\right), 
\end{equation}
for $a_i = 1, \ldots, k_i$ and $ i = 1, \ldots, s$.
For the first component, $i = 1$, since $k_{i-1} = 0$, 
the left hand side of \eqref{eqn:QK-rel} is simply $(-)^{k_i -1} q_i \left(X_{a_i}^{(i)}\right)^{k_i}$.

\subsection{Characteristic polynomials and symmetrization}
\label{sect:charpoly}

Next we will symmetrize \eqref{eqn:QK-rel} 
to obtain 
characteristic polynomials and then use Vieta relations 
to form the physics relations for the quantum K-theory ring,
much as we did for Grassmannians in section~\ref{sect:gr:symm}.

First, after some rearrangement, \eqref{eqn:QK-rel} becomes
\begin{eqnarray}
\lefteqn{
    (-)^{k_i - 1} e_{k_{i+1}}\left( X^{(i+1)} \right) q_i 
\left(X^{(i)}_{a_i}\right)^{k_i - k_{i-1}} 
\prod_{b_{i-1} =1}^{k_{i-1}}\left(X^{(i)}_{a_i} - X^{(i-1)}_{b_{i-1}}\right) 
} \nonumber \\
& \hspace*{1.0in} = &
 (-)^{k_{i+1}} e_{k_i} \left( X^{(i)} \right)
 \prod_{b_{i+1} = 1}^{k_{i+1}} \left(X^{(i)}_{a_i} - X^{(i+1)}_{b_{i+1}}\right)
\label{eq:char:int1}
\end{eqnarray}
for $a = 1, \cdots, k_i$,
where $e_i(x)$ is the $i$-th elementary symmetric polynomial in 
the indeterminates $\{x_a\}$.

Now we use the expansion
\begin{equation}
    \prod_{j = 1}^n (\xi - x_j) = \xi^n - e_1(x) \xi^{n-1} + e_2(x) \xi^{n-2} + \cdots + (-)^n e_n(x),
\end{equation}
to write~(\ref{eq:char:int1}) as
\begin{equation}
\begin{aligned}
    (-)^{k_i - 1} e_{k_{i+1}} \left(X^{(i+1)} \right) q_i 
\left[\xi^{k_i} 
 - e_1\left(X^{(i-1)}\right) \xi^{k_i - 1} + \cdots
 + (-)^{k_{i-1}} e_{k_{i-1}}\left(X^{(i-1)}\right) \xi^{k_i - k_{i-1}}\right]\cr
    = \:
(-)^{k_{i+1}} e_{k_i}\left(X^{(i)}\right)\left[
\xi^{k_{i+1}} 
- e_1\left(X^{(i+1)}\right) \xi^{k_{i+1} - 1} + \cdots
 + (-)^{k_{i+1}} e_{k_{i+1}}\left(X^{(i+1)}\right)\right].
\end{aligned}
\end{equation}
for 
\begin{equation}
\xi \: = \: X^{(i)}_{a_i},
\end{equation}
or equivalently, after rearrangement,
\begin{equation}
    \sum_{\ell = 0}^{k_{i+1}} (-)^{\ell} \xi^{k_{i+1} - \ell}
 \left[e_{k_i} \left(X^{(i)}\right) e_{\ell}\left(X^{(i+1)}\right)
 + q_i e_{k_{i+1}}\left(X^{(i+1)}\right) e_{\ell - k_{i+1} +k_i}\left(X^{(i-1)}\right)\right]
\: = \: 0,
\end{equation}
in conventions for which $e_{\ell}(x) = 0$ for $\ell < 0$.
We call this the characteristic polynomial equation, 
which is of order $k_{i+1}$. 
We denote the $k_{i+1}$ roots of this equation by $w$'s.
Of those $k_{i+1}$ roots, $k_i$ are the $X^{(i)}_a$,
and the remainder are denoted $\bar{X}^{(i)}_a$.
Then from Vieta's formula~(\ref{eq:vieta})
we have the quantum K-theory ring relations,
\begin{equation}\label{eqn:rel}
    e_{k_i} \left(X^{(i)}\right) e_{\ell}(w) 
\: =  \:
e_{k_i}\left(X^{(i)}\right) e_{\ell}\left(X^{(i+1)}\right) 
\: + \:
 q_i e_{k_{i+1}}\left(X^{(i+1)}\right) e_{\ell-k_{i+1}+ k_i}\left(X^{(i-1)}\right),
\end{equation}
for $\ell = 0, \cdots, k_{i+1}$, where
\begin{equation}  \label{eq:relsum}
    e_{\ell}(w) = \sum_{r = 0}^{k_{i+1} - k_i} e_{\ell-r}\left(X^{(i)}\right) 
e_r\left(\bar{X}^{(i)}\right).
\end{equation}

This expression can be simplified.  First  note that
for $\ell < k_{i+1} - k_i$ or $\ell > k_{i+1}-k_i + k_{i-1}$,
from~(\ref{eqn:rel}) we have
\begin{equation} \label{eq:w=z}
    e_{\ell}(w) = e_{\ell}\left(X^{(i+1)}\right),
\end{equation}
where in our notation, $e_{\ell}\left(X^{(i-1)}\right) = 0$ if $\ell < 0$ 
or $\ell > k_{i-1}$
(since there are $k_{i-1}$ $X^{(i-1)}$'s).

When $\ell = k_{i+1} - k_i$, equation~(\ref{eqn:rel}) implies
\begin{equation} \label{eq:ekint}
    e_{k_i}\left(X^{(i)}\right) e_{k_{i+1} - k_i}(w) 
\: = \:
 e_{k_i}\left(X^{(i)}\right) e_{k_{i+1} - k_i}\left(X^{(i+1)}\right)
\: + \:
 q_i \, e_{k_{i+1}}\left(X^{(i+1)}\right).
\end{equation}
To eliminate $e_{k_{i+1}}\left(X^{(i+1)}\right)$ in this equation,
we take $\ell = k_{i+1}$ in \eqref{eq:relsum}
which, using~(\ref{eq:w=z}), implies
\begin{equation}  \label{eq:326}
 e_{k_i}\left(X^{(i)}\right) e_{k_{i+1} - k_i}\left(\bar{X}^{(i)}\right)
\: = \:
 e_{k_{i+1}}\left(X^{(i+1)}\right)
\end{equation}
(as the only nonzero contribution to the sum is from the case
$s = k_{i+1} - k_i$).
 Therefore, plugging this into~(\ref{eq:ekint}),
\begin{equation}
    e_{k_{i+1} - k_i}(w)
\: = \:
 e_{k_{i+1} - k_i}\left(X^{(i+1)}\right)
\: + \:
 q_i \, e_{k_{i+1} - k_i}\left(\bar{X}^{(i)}\right).
\end{equation}
Applying~(\ref{eq:relsum}), we can write this as
\begin{equation}  \label{eq:int2}
\sum_{r=0}^{k_{i+1}-k_i-1} e_{k_{i+1}-k_i-r}\left(X^{(i)}\right) 
e_r\left(\overline{X}^{(i)}\right)
\: + \: (1 - q_i) \,
e_{k_{i+1}-k_i}\left(\overline{X}^{(i)}\right)
\: = \: e_{k_{i+1} - k_i}\left(X^{(i+1)}\right).
\end{equation}

Now, we are ready to simplify and derive the quantum K theory relations.
Using~(\ref{eq:326}) to simplify~(\ref{eqn:rel}), we have
\begin{equation}
e_{\ell}(w) \: = \: e_{\ell}\left(X^{(i+1)}\right) \: + \:
q_i \, e_{k_{i+1}-k_i}\left(\overline{X}^{(i)}\right) 
e_{\ell - k_{k+1}+k_i}\left(X^{(i-1)}\right).
\end{equation}
Using~(\ref{eq:relsum}), this becomes
\begin{equation}  \label{eq:keyreln}
\sum_{r=0}^{k_{i+1}-k_i} e_{\ell-r}\left(X^{(i)}\right) 
e_r\left(\overline{X}^{(i)}\right)
\: = \:
e_{\ell}\left(X^{(i+1)}\right) \: + \:
q_i \, e_{k_{i+1}-k_i}\left(\overline{X}^{(i)}\right) 
e_{\ell - k_{i+1}+k_i}\left(X^{(i-1)}\right).
\end{equation}
This is the the proposed quantum K theory relation, and is also
the key algebraic relation needed to relate to 
$\lambda_y$ class presentations in the next section.

Proceeding as for Grassmannians, let us solve this equation
algebraically for the $e_{\ell}(\overline{X}^{(i)})$.
To do so, we rewrite the expression above as the degree-$\ell$ part
of 
\begin{eqnarray}
\lefteqn{
\left( \sum_{r=0}^{k_{i}} y^r e_r\left( X^{(i)} \right) \right)
\left( \sum_{t=0}^{k_{i+1}-k_i} y^t e_t\left( \overline{X}^{(i)} \right) 
\right)
}  \\
& = &
\left( \sum_{r=0}^{k_{i+1}} y^r e_r\left( X^{(i+1)} \right) \right)
\: + \:
q_i \, y^{k_{i+1}-k_i} \, e_{k_{i+1}-k_i}\left(\overline{X}^{(i)}\right) 
\left( \sum_{r=0}^{k_{i-1}} y^r e_r\left( X^{(i-1)} \right) \right),
\nonumber
\end{eqnarray}
hence
\begin{eqnarray}
\left( \sum_{r=0}^{k_{i+1}-k_i} y^r e_r\left( \overline{X}^{(i)} \right) 
\right)
& = &
\left( \sum_{r=0}^{k_{i+1}} y^r e_r\left( X^{(i+1)} \right) \right)
\left( \sum_{t=0}^{\infty} (-)^t y^t h_t\left( X^{(i)} \right) \right)
\\
& &
\: + \:
q_i \, y^{k_{i+1}-k_i} \, e_{k_{i+1}-k_i}\left(\overline{X}^{(i)}\right) 
\left( \sum_{r=0}^{k_{i-1}} y^r e_r\left( X^{(i-1)} \right) \right)
\left( \sum_{t=0}^{\infty} (-)^t y^t h_t\left( X^{(i)} \right) \right).
\nonumber
\end{eqnarray}

To simplify the expression above, we define $\{ \hat{X}^{(i)} \}$ by
\begin{equation}  \label{eq:flag:hatX:defn}
e_{\ell}\left( \hat{X}^{(i)} \right) \: = \:
\sum_{r=0}^{k_{i+1}} (-)^r e_{\ell-r}\left( X^{(i+1)} \right) h_r\left(
X^{(i)} \right),
\end{equation}
then we see that for $\ell < k_{i+1}-k_i$,
\begin{equation}
e_{\ell}\left( \overline{X}^{(i)} \right)
\: = \:
e_{\ell}\left( \hat{X}^{(i)} \right),
\end{equation}
and for $\ell = k_{i+1}-k_i$,
\begin{equation}
e_{k_{i+1}-k_i}\left( \overline{X}^{(i)} \right)
\: = \:
e_{k_{i+1}-k_i}\left( \hat{X}^{(i)} \right)
\: + \: q_i \,  e_{k_{i+1}-k_i}\left(\overline{X}^{(i)}\right) 
e_{0}\left( \hat{X}^{(i-1)} \right),
\end{equation}
hence
\begin{equation}
e_{k_{i+1}-k_i}\left( \overline{X}^{(i)} \right)
\: = \:
(1-q_i)^{-1} e_{k_{i+1}-k_i} \left( \hat{X}^{(i)} \right).
\end{equation}

In summary,
\begin{equation}
e_{\ell}\left(\overline{X}^{(i)} \right)
\: = \: 
\left\{ \begin{array}{cl}
e_{\ell}\left( \hat{X}^{(i)} \right) & \ell < k_{i+1}-k_i,
\\
(1-q_i)^{-1} e_{k_{i+1}-k_i}\left( \hat{X}^{(i)} \right) &
\ell = k_{i+1}-k_i.
\end{array} \right.
\end{equation}

\subsection{Interpretation in terms of bundles: $\lambda_y$ class presentation}
\label{sect:lambda-pres}

In this section we interpret~(\ref{eq:keyreln}) in terms of bundles,
and utilize it to generate the $\lambda_y$ class presentations of the
quantum K theory ring.  Although the details are more elaborate,
the underlying logic is the same as that presented for
Grassmannians in section~\ref{sect:gr:lambda-pres}.
(Also, just as there, we emphasize that
any interpretation
may be slightly ambiguous; we will utilize comparisons to mathematics to
justify our proposal.)

For simplicity, we focus on three successive steps $k_{i-1}, k_i, k_{i+1}$
in the flag of
a flag manifold $F(k_1, k_2, \cdots, k_s, n)$.
These correspond to three bundles, ${\cal S}_{i-1} \subset {\cal S}_i \subset {\cal S}_{i+1}$,
of ranks $k_{i-1}, k_i, k_{i+1}$.
We denote the quotient bundle by ${\cal S}_{i+1} / {\cal S}_i$, 
which is of rank $k_{i+1} - k_i$.
The $\{ X^{(i)} \}$ are associated with the Chern roots of the universal
subbundle ${\cal S}_i$.  In addition, it is natural to associate the
$\{ \hat{X}^{(i)} \}$ with the Chern roots of the universal quotient bundle
${\cal S}_{i+1}/{\cal S}_i$.  Classically, this follows from the
defining property~(\ref{eq:flag:hatX:defn}) of the 
$\{ \hat{X}^{(i)} \}$.  In more detail, this is a consequence of the
short exact sequence
\begin{equation}
0 \: \longrightarrow \: {\cal S}_i \: \longrightarrow \:
{\cal S}_{i+1} \: \longrightarrow \: {\cal S}_{i+1}/{\cal S}_i
\: \longrightarrow \: 0,
\end{equation}
which implies
\begin{equation}
c\left( {\cal S}_i \right) \,
c\left( {\cal S}_{i+1}/{\cal S}_i \right) \: = \:
c\left( {\cal S}_{i+1} \right),
\end{equation}
and hence the relation~(\ref{eq:flag:hatX:defn}), after algebra.
(This association will be justified by the fact that this will correctly
reproduce rigorous results for quantum K theory rings of e.g.~incidence
varieties and also quantum cohomology rings.)

We formally associate the $\{\bar{X}^{(i)} \}$
 with a bundle $\widetilde{\mathcal R}_i$ of rank $k_{i+1} - k_i$.
From the results of the last section, we can identify
\begin{equation}  \label{eq:dictionary}
e_{\ell}\left(\overline{X}^{(i)}\right) \: \leftrightarrow \:
\wedge^{\ell} \widetilde{\mathcal R}_i \: = \:
\left\{ \begin{array}{cl}
\wedge^{\ell}  ({\cal S}_{i+1}/{\cal S}_i) & \ell < k_{i+1} - k_i,
\\
(1-q_i)^{-1} \,  \det  ({\cal S}_{i+1}/{\cal S}_i)& \ell = k_{i+1} - k_i.
\end{array}
\right.
\end{equation}
The reader should note that this means that although for $q_i = 0$,
$\widetilde{\mathcal R}_i = {\cal S}_{i+1}/{\cal S}_i$, for nonzero $q_i$, 
$\widetilde{\mathcal R}_i$ is
not a classical bundle, but instead appears to be more nearly some sort of
quantum exterior product (which we shall not try to define here, aside
from the statement above).

Now, we are ready to interpret the (symmetrized)
quantum K theory relation~(\ref{eq:keyreln}), namely
\begin{equation}  
\sum_{r=0}^{k_{i+1}-k_i} e_{\ell-r}\left(X^{(i)}\right) 
e_r\left(\overline{X}^{(i)}\right)
\: = \:
e_{\ell}\left(X^{(i+1)}\right) \: + \:
q_i e_{k_{i+1}-k_i}\left(\overline{X}^{(i)}\right) 
e_{\ell - k_{i+1}+k_i}\left(X^{(i-1)}\right).
\end{equation}
Interpreting ordinary products in the algebraic relation above as
quantum products $\star$, and also using the dictionary derived above for
$e_{\ell}(\overline{X})$, we have
\begin{equation}
\sum_{r=0}^{k_{i+1}-k_i} \wedge^{\ell-r}( {\cal S}_i ) \star
\wedge^r( \widetilde{\mathcal R}_i) \: = \: 
\wedge^{\ell}( {\cal S}_{i+1} ) \: + \:
q_i \left( \det \widetilde{\mathcal R}_i \right) \star
\wedge^{\ell - k_{i+1} + k_i}( {\cal S}_{i-1} ),
\end{equation}
or more elegantly,
\begin{equation}  \label{eq:qk:flag:reln:lambda0}
\lambda_y( {\cal S}_i) \star
\lambda_y( \widetilde{\mathcal R}_i ) \: = \:
\lambda_y({\cal S}_{i+1}) \: + \:
q_i y^{k_{i+1}-k_i} \left( \det \widetilde{\mathcal R}_i \right) \star
\lambda_y( {\cal S}_{i-1} ).
\end{equation}
Equation~(\ref{eq:qk:flag:reln:lambda0}) is our proposal for the
relation defining the $T$-equivariant quantum K theory ring of a general
partial flag manifold.

Alternatively, we can write this without using the $\widetilde{\mathcal R}_i$.
Returning to the relation~(\ref{eq:keyreln}) and using the
dictionary~(\ref{eq:dictionary}), we have
\begin{eqnarray}
\lefteqn{
\sum_{r=0}^{k_{i+1}-k_i-1} \wedge^{\ell-r} {\cal S}_i \, \star \, \wedge^r  ({\cal S}_{i+1}/{\cal S}_i)
\: + \: \frac{1}{1-q_i} \det  ({\cal S}_{i+1}/{\cal S}_i)\,  \star \, \wedge^{\ell - k_{i+1} + k_i}
{\cal S}_i
} \nonumber \\
&\hspace*{0.75in} = & 
\wedge^{\ell} {\cal S}_{i+1} \: + \:
\frac{q_i}{1-q_i} \det ({\cal S}_{i+1}/{\cal S}_i) \star \wedge^{\ell- k_{i+1} + k_i} {\cal S}_{i-1},
\end{eqnarray}
which after a little algebra implies
\begin{eqnarray}  \label{eq:qk:flag:reln}
\lefteqn{
\sum_{r=0}^{k_{i+1}-k_i} \wedge^{\ell-r} {\cal S}_i \, \star \, \wedge^r ({\cal S}_{i+1}/{\cal S}_i)
} \\
& \hspace*{0.5in} = & \wedge^{\ell} {\cal S}_{i+1} \: - \:
\frac{q_i}{1-q_i} \det  ({\cal S}_{i+1}/{\cal S}_i) \star \left( \wedge^{\ell- k_{i+1} + k_i} {\cal S}_{i}
\: - \:
 \wedge^{\ell - k_{i+1} + k_i}
{\cal S}_{i-1} \right).
\nonumber
\end{eqnarray}
By multiplying factors of $y^{\ell}$, we can rewrite it in terms of
$\lambda_y$ classes, as
\begin{equation}  \label{eq:qk:flag:reln:lambda}
\lambda_y({\cal S}_i) \star \lambda_y({\cal S}_{i+1}/{\cal S}_i) \: = \:
\lambda_y({\cal S}_{i+1}) \: - \:
y ^{k_{i+1} - k_{i}} \frac{q_i}{1-q_i} \det ({\cal S}_{i+1}/{\cal S}_i) \star
\left( \lambda_y({\cal S}_{i}) - \lambda_y({\cal S}_{i-1}) \right).
\end{equation}
This is a second form of
our claimed presentation of the quantum K theory relations
in terms of $\lambda_y$ classes.
We remind the reader that in our conventions,
$k_0 = 0$, $k_{s+1} = n$,
${\cal S}_0 = 0$, ${\cal S}_{s+1} = {\mathbb C}^n$, and
the equivariant structure is encoded implicitly in
${\mathbb C}^n$.

As a consistency check of~(\ref{eq:qk:flag:reln:lambda}), 
the reader may note that this immediately
reduces to the result \cite[theorem 1.1]{Gu:2022yvj} for the
$T$-equivariant quantum K theory ring of the Grassmannian
$G(k,n)$, namely
\begin{equation}\label{eq:lambdayGr} 
\lambda_y({\cal S}) \star \lambda_y(\mathcal Q) \: = \:
\lambda_y({\mathbb C}^n) - y^{n-k} \frac{q}{1-q} \det {\mathcal Q} \star
\left( \lambda_y({\cal S}) - 1 \right).
\end{equation}

As another consistency check of~(\ref{eq:qk:flag:reln:lambda}), 
we observe that the classical
($q_i \rightarrow 0$)
limit can be immediately derived from the short exact sequence
\begin{equation}
0 \: \longrightarrow \: {\cal S}_i \: \longrightarrow \: {\cal S}_{i+1} \:
\longrightarrow \: {\cal S}_{i+1}/{\cal S}_i \: \longrightarrow \: 0.
\end{equation}

\begin{remark}
The equation \eqref{eq:qk:flag:reln:lambda}
suggests the following interpretation.
Realize the partial flag manifold $F(k_1, \ldots, k_s,n)$ as the Grassmann
bundle $Gr(k_i-k_{i-1}, {\cal S}_{i+1}/{\cal S}_{i-1}) \to 
F(k_1, \ldots, k_{i-1}, k_{i+1}, \ldots, k_s,n)$. This is equipped with the
tautological sequence 
\begin{equation*}
0 \: \longrightarrow \: {\cal S}_i / {\cal S}_{i-1} \: \longrightarrow \:
{\cal S}_{i+1}/{\cal S}_{i-1} \: \longrightarrow \: {\cal S}_{i+1}/{\cal S}_i
\: \longrightarrow \: 0.
\end{equation*}
We may formally divide \eqref{eq:qk:flag:reln:lambda}
by $\lambda_y(\mathcal{S}_{i-1})$ to obtain
\begin{equation}\label{E:verticalQK}  
\frac{\lambda_y(\mathcal{S}_{i})}{\lambda_y(\mathcal{S}_{i-1})} \star \lambda_y(\mathcal{S}_{i+1}/\mathcal{S}_{i}) =\frac{\lambda_y(\mathcal{S}_{i+1})}{\lambda_y(\mathcal{S}_{i-1})}
- y^{k_{i+1} - k_i}\frac{q_i}{1-q_i} \det (\mathcal{S}_{i+1}/\mathcal{S}_i) \star
 \left(\frac{\lambda_y(\mathcal{S}_{i})}{\lambda_y(\mathcal{S}_{i-1})} -1\right) \/.
 \end{equation}
This holds in the classical K theory ring (i.e., when $q_i=0$), since 
 $\lambda_y( {\cal S}_i/{\cal S}_{i-1})=\frac{\lambda_y({\cal S}_i) }{
\lambda_y({\cal S}_{i-1}) }$ and 
$\lambda_y( {\cal S}_{i+1}/{\cal S}_{i-1})=\frac{\lambda_y({\cal S}_{i+1}) }{
\lambda_y({\cal S}_{i-1}) }$, and because of the usual K-theoretic Whitney relations. 
The identity \eqref{E:verticalQK} may be interpreted as a relative version of the 
quantum relations \eqref{eq:lambdayGr} on the Grassmannian 
 $G(k_i-k_{i-1}, \mathbb{C}^{k_{i+1}-k_{i-1}}$). 

\end{remark}

For later use, it may be helpful to specialize the 
relation~(\ref{eq:qk:flag:reln}).
First, for the case 
$\ell = k_{i+1}$, equation~(\ref{eq:qk:flag:reln}) reduces
to
\begin{equation}\label{eqn:rel2}
    \det {\cal S}_i \star \det ({\cal S}_{i+1}/{\cal S}_i) 
\: =  \:
(1-q_i)\det {\cal S}_{i+1}.
\end{equation}
Similarly,
by multiplying factors of $\det {\cal S}_i$ for $\ell < k_{i+1}$
and using the relation above, one gets
\begin{eqnarray}  
\lefteqn{
    \left(\wedge^{k_{i+1}} {\cal S}_{i+1}\right) \star \left( \wedge^{\ell-k_{i+1} + k_i} {\cal S}_i - q_i \wedge^{\ell-k_{i+1}+k_i} {\cal S}_{i-1}\right) 
} \nonumber \\
& \hspace*{0.5in} = & 
     \left(\wedge^{k_i} {\cal S}_i\right) \star \left[\wedge^{\ell} {\cal S}_{i+1} - \sum_{s = 0}^{k_{i+1} - k_i - 1} \wedge^{\ell-s} {\cal S}_i \star \wedge^s ({\cal S}_{i+1}/{\cal S}_i)\right].
\label{eqn:rel1}
\end{eqnarray}

An equivalent formulation  of the $\lambda_y$ class relations is in terms of
corresponding K-theoretic Chern roots.  We can write equation~(\ref{eq:qk:flag:reln:lambda})
as 
\begin{eqnarray}
\lefteqn{
\left[ \prod_{j=1}^{k_i} \left( 1 + y x_j \right) \right] 
\cdot
\left[ \prod_{j=1}^{k_{i+1} - k_i} \left( 1 + y v_j \right) \right]
}  \\
& = &
\prod_{j=1}^{k_{i+1}} \left( 1 + y z_j \right) \: - \:
\frac{q_i}{1-q_i} y^{k_{i+1} - k_i} 
\left[ \prod_{j=1}^{k_{i+1}-k_i} v_j \right]
\left[ \prod_{j=1}^{k_i} \left( 1 + y x_j \right)
\: - \:
\prod_{j=1}^{k_{i-1}} \left( 1 + y u_j \right) \right],
\nonumber
\end{eqnarray}
where
\begin{equation}
x \sim {\cal S}_i, \: \: \:
v \sim {\cal S}_{i+1}/{\cal S}_i, \: \: \:
z \sim {\cal S}_{i+1}, \: \: \:
u \sim {\cal S}_{i-1}.
\end{equation}

\subsection{Shifted variables}
\label{sect:shifted}

We can also use the shifted Wilson line basis discussed in 
\cite{Gu:2020zpg,Gu:2022yvj}. 
The shifted variables $z$ are defined by $z_a \equiv 1- X_a$.
Defining
\begin{equation}
    c^{(i+1)} = \prod_{b_{i+1} = 1}^{k_{i+1}} \left(1 - z^{(i+1)}_{b_{i+1}}\right),
\end{equation}
we can rewrite~(\ref{eqn:QK-rel}) in terms of shifted variables as
\begin{equation}
    (-)^{k_i - 1} q_i \left(1 - z^{(i)}_{a_i}\right)^{k_i - k_{i-1}-1}c^{(i+1)} \prod_{b_{i-1} = 1}^{k_{i-1}} \left(z_{b_{i-1}}^{(i-1)} - z^{(i)}_{a_i}\right)
    \: =  \:
    \left(\prod_{b_i \neq a_i} \left(1 - z_{b_i}^{(i)}\right)\right)\prod_{b_{i+1} = 1}^{k_{i+1}} \left(z^{(i)}_{a_i} - z_{b_{i+1}}^{(i+1)}\right).
\end{equation}
Following the same argument in \cite{Gu:2022yvj}, we may rewrite this in the form
\begin{equation}
    \left(z_{a_i}^{(i)}\right)^{k_{i+1}} 
+ 
\sum_{r = 0}^{k_{i+1} - 1}(-1)^{k_{i+1} - r} \left(z_{a_i}^{(i)}\right)^r 
g^{(i)}_{n-r}\left(z^{(i-1)}, z^{(i)}, z^{(i+1)}, q_i\right).
\end{equation}
where $g^{(i)}_r$ are symmetric polynomials in $z^{(i)}_1, \dots, z^{(i)}_{k_i}$ and $z^{(i+1)}_1, \dots, z^{(i+1)}_{k_{i+1}}$. To state the formula for the this polynomial, 
we make the following definitions borrowed from \cite{Gu:2022yvj}. Set
\begin{eqnarray}
    c^{(i)} & = & \prod_{a_i = 1}^{k_i} \left(1 - z_{a_i}^{(i)}\right) = \sum_{s \ge 0} (-1)^s e_s\left(z^{(i)}\right),
\\
    c^{(i)}_{\le j} & = & \sum_{r = 0}^j (-1)^r e_r\left(z^{(i)}\right),
\\
    c^{(i)}_{\ge j} & = & (-1)^j \left(c^{(i)} - c^{(i)}_{\le j-1}\right).
\end{eqnarray}
Similarly, one defines $c^{(i+1)}, c^{(i+1)}_{\le j}, c^{(i+1)}_{\ge j}$. Set
\begin{eqnarray}
\lefteqn{
    c'_{\ge \ell}(z^{(i)}, z^{(i+1)})
}  \\
& = & e_\ell\left(z^{(i+1)}\right) + e_{\ell - 1}\left(z^{(i+1)}\right) c^{(i)}_{\ge 2} + e_{\ell -2}\left(z^{(i+1)}\right) c^{(i)}_{\ge 3} + \dots + e_{\ell - k_i + 1}(z^{(i+1)}) c^{(i)}_{\ge k_i}
\nonumber
\end{eqnarray}
for $k_i \ge 2$ and $c'_{\ge \ell}(z^{(i)}, z^{(i+1)})  = e_\ell(z^{(i+1)})$ when $k_i = 1$.

Define the matrices
\begin{equation}
\begin{gathered}
    E=\left(\begin{array}{cccc}
    -1 & 0 & \ldots & 0 \\
    -e_1(z^{(i)}) & -1 & \ldots & 0 \\
    \vdots & \vdots & \ddots & 0 \\
    -e_{k_i-1}(z^{(i)}) & -e_{k_i-2}(z^{(i)}) & \ldots & -1
    \end{array}\right) ; \quad
    C_{\geq k_{i+1}-k_i+2}^{(i+1)}=\left(\begin{array}{c}
    c_{\geq k_{i+1}-k_i+2}^{(i+1)} \\
    \vdots \\
    c_{\geq k_{i+1}}^{(i+1)} \\
    0
    \end{array}\right).
    \end{gathered}
\end{equation}
Then the polynomial coefficients $g_\ell^{(i)}$ are given by
\begin{equation}
    g_\ell^{(i)} = \begin{cases}
        c'_{\ge \ell} \left(z^{(i)}, z^{(i+1)}\right) & 1 \le \ell \le k_{i+1} - k_i\\
        c'_{\ge \ell} \left(z^{(i)}, z^{(i+1)}\right) + \left(E \cdot C^{(i+1)}_{\ge k_{i+1} - k_i +2}\right)_\ell + (-1)^{k_{i+1} + k_i} q_i c^{(i+1)} \alpha_\ell^{(i)} & k_{i+1} - k_i + 1 \le \ell \le k_{i+1}
    \end{cases}
\end{equation}
where
\begin{equation}
    \alpha_\ell^{(i)} = \sum_{r = 0}^{k_{i-1}} \binom{k_i - k_{i-1} - 1}{k_{i+1} - k_{i-1} + r - \ell} e_r\left(z^{(i-1)}\right).
\end{equation}
Define a characteristic polynomial $f^{(i)}\left(\xi, z^{(i-1)}, z^{(i)}, z^{(i+1)}, q_i\right)$ by
\begin{equation}\label{eqn:characteristic-eqn-shifted}
    f^{(i)}\left(\xi, z^{(i-1)}, z^{(i)}, z^{(i+1)}, q_i\right)
 = \xi^{k_{i+1}}
 + \sum_{r=0}^{k_{i+1} - 1} (-1)^{k_{i+1} - r} \xi^r g_{k_{i+1} - r}^{(i)}\left(z^{(i-1)}, z^i, z^{(i+1)}, q_i\right).
\end{equation}
We have $f^{(i)} = 0$ whenever $\xi = z^{(i)}_{a_i}$ for $a_i = 1, \dots, k_i$. Since $f^{(i)}$ is a degree $k_{i+1}$ polynomial in $\xi$, the equation $f^{(i)} = 0$ has $k_{i+1}$ roots, which include all $z_{a_i}^{(i)}$'s. Let $\{z^{(i)}, \hat{z}^{(i)}\} = \{z_1^{(i)}, \cdots, z^{(i)}_{k_i}; \hat{z}^{(i)}_{k_i + 1}, \cdots, \hat{z}^{(i)}_{k_{i+1}}\}$ denote the $k_{i+1}$ roots of \eqref{eqn:characteristic-eqn-shifted}.
From Vieta's formula, we then have the relations
\begin{equation}\label{eqn:relation-shifted}
    \sum_{r+t = \ell} e_r\left(z^{(i)}\right) e_t\left(\hat{z}^{(i)}\right) = g_\ell^{(i)} \left(z^{(i-1)}, z^{(i)}, z^{(i+1)}, q_i\right).
\end{equation}
We conjecture that these relations define the $T$-equivariant quantum
K theory ring of the partial flag manifold, generalizing the
corresponding presentation for Grassmannians 
(denoted there the ``Coulomb branch presentation'') given in
\cite[section 10]{Gu:2022yvj}.

\subsection{Consistency test: quantum cohomology}
\label{sect:cons:qh}

In this section, we take the two-dimensional limit of the quantum K theory,
to recover a prediction for $T$-equivariant
quantum cohomology $QH_T^*$, 
which can be checked against results in the literature, see for example
\cite{Astashkevich:1993ks,cf1,Donagi:2007hi,gk}.

To that end, we take the theory to be defined on a 3-manifold
$S^{1}\times\Sigma$ for some Riemann surface $\Sigma$,
where $S^{1}$ has diameter $L$.  Then, as discussed
in e.g.~\cite{Gu:2020zpg}, the quantities appearing in this section can be
expanded as follows:
\begin{eqnarray*}
  q_{i}=L^{k_{i+1}-k_{i-1}}q_{i,2d}, 
&& z=1-X=-L\sigma-\frac{L^{2}}{2}\sigma^{2}-\cdots \\
\hat{z} 
&=&-L\hat{\sigma}-
\cdots\\
   T_{i}=\exp(Lt_{i})&=&1+Lt_{i}+\frac{L^{2}}{2}t_{i}^{2}+\cdots . 
\end{eqnarray*}
We will first compute directly by expanding the $\lambda_y$ relations,
then we will separately back up to the twisted effective superpotential,
and repeat the same computation there.

\subsubsection{Expansion of the $\lambda_{y}$ class}
\label{sect:qh:exp-lambda}

In section~\ref{sect:shifted}, we gave results for quantum K theory relations
in terms of shifted variables.  In this section we describe their two-dimensinoal
limits.  To that end, we first observe that in the limit described above,
\begin{eqnarray*}
  c^{(i)}_{\geq j} &\mapsto& (-L)^{j}e_{j}(\sigma^{(i)})+{\cal O}(L^{j}),
\\
   c^{\prime}_{\geq\ell}\left(z^{(i)},z^{(i+1)}\right) &\mapsto& (-L)^{\ell}e_{\ell}(\sigma^{(i+1)})+{\cal O}(L^{\ell}),
\\
   \left(E\cdot C\right)_{\ell}&\mapsto& {\cal O}(L^{\ell}),
\end{eqnarray*}
where $e_{0}(\sigma^{(i)})=1$.
The first nonzero coefficient in
\begin{equation}
    \alpha_\ell^{(i)} \: = \: \sum_{r = 0}^{k_{i-1}} \binom{k_i - k_{i-1} - 1}{k_{i+1} - k_{i-1} + r - \ell} e_{r}\left(z^{(i-1)}\right)
\end{equation}
is at
\begin{equation}
  r \: = \: \ell+k_{i-1}-k_{i+1}.
\end{equation}
As a result, the two-dimensional limit of $\alpha_\ell^{(i)}$ is
\begin{equation}
  \alpha_\ell^{(i)}\mapsto (-L)^{\ell+k_{i-1}-k_{i+1}}e_{\ell+k_{i-1}-k_{i+1}}\left(\sigma^{(i-1)}\right)+{\cal O}(L^{\ell+k_{i-1}-k_{i+1}}).
\end{equation}

Using the computations above, the two-dimensional
limit of the polynomial coefficients $g_\ell^{(i)}$ is given by
\begin{eqnarray} \label{COF}
\lefteqn{
    g_{\ell,2d}^{(i)} 
} \\
& = & \begin{cases}
        (-L)^{\ell}e_{\ell} \left(\sigma^{(i+1)}\right)+{\cal O}(L^{\ell}) & 1 \le \ell \le k_{i+1} - k_i\\
        (-L)^{\ell}e_{\ell} \left(\sigma^{(i+1)}\right) + (-1)^{\ell+k_i+k_{i-1}} q_{i,2d}\cdot e_{\ell+k_{i-1}-k_{i+1}}\left(\sigma^{(i-1)}\right)+{\cal O}(L^{\ell}) & k_{i+1} - k_i < \ell \le k_{i+1}.
    \end{cases}
\nonumber
\end{eqnarray}
Finally, we have
\begin{equation}
   \sum_{r+t = \ell} e_s\left(z^{(i)}\right) e_t\left(\hat{z}^{(i)}\right)
\: \mapsto \:
 (-L)^{\ell} \sum_{r+t = \ell}e_r\left(\sigma^{(i)}\right) e_t\left(\hat{\sigma}^{(i)}\right).
\end{equation}
Applying the computations above to~(\ref{eqn:relation-shifted}),
this gives the quantum cohomology ring relations
\begin{eqnarray} \label{Vieta}
\lefteqn{
    \sum_{s+t = \ell}
e_s\left(\sigma^{(i)}\right) e_t\left(\hat{\sigma}^{(i)}\right) 
}
\\
& = & \begin{cases}
        e_{\ell} \left(\sigma^{(i+1)}\right) & 1 \le \ell \le k_{i+1} - k_i, \\
        e_{\ell} \left(\sigma^{(i+1)}\right) + (-1)^{k_i-k_{i-1}} q_{i,2d}\cdot e_{\ell+k_{i-1}-k_{i+1}}\left(\sigma^{(i-1)}\right) & k_{i+1} - k_i + 1 \le \ell \le k_{i+1},
    \end{cases}
\nonumber
\end{eqnarray}
which can be written more succinctly in terms of $T$-equivariant
total Chern classes $c^T$ as follows:
\begin{equation}  \label{eq:genl-qh}
c^T({\cal S}_i) \star c^T\left( {\cal S}_{i+1}/{\cal S}_i \right) \: = \:
c^T\left( {\cal S}_{i+1} \right) \: + \:
(-)^{ k_i - k_{i-1} } q_i c^T\left( {\cal S}_{i-1} \right).
\end{equation}

\subsubsection{Expansion of the twisted effective superpotential}

As an alternative procedure, we will check in this subsection that
we can also get the same result from
the twisted effective superpotential.
To that end, it can be shown that
\begin{equation}
  \widetilde{W}_{3d}\mapsto L\widetilde{W}_{2d}+{\cal O}(L) ,
\end{equation}
which implies
\begin{eqnarray}
  \widetilde{W}_{2d}
& = &
\sum^{N}_{i=1}\sum^{k_{i}}_{a=1}\Sigma^{(i)}_{a}
\Biggl[ -t_{i} \: + \: i\pi\left(k_{i}-1\right)
\: - \: \sum^{k_{i+1}}_{r=1}\left(\log\left(\Sigma^{(i)}_{a}-\Sigma^{(i+1)}_{r}\right)-1\right)
\nonumber \\
& & \hspace*{1.5in}
\: + \: \sum^{k_{i-1}}_{r=1}\left(\log\left(\Sigma^{(i-1)}_{r}-\Sigma^{(i)}_{a}\right)-1\right)\Biggr].
\end{eqnarray}
The vacuum equations are
\begin{equation*}
  \prod^{k_{i+1}}_{r=1}\left(\sigma^{(i)}_{a}-\sigma^{(i+1)}_{r}\right)
\: = \:
(-1)^{k_{i}-1}q_{i,2d}\prod^{k_{i-1}}_{r=1}\left(\sigma^{(i-1)}_{r}-\sigma^{(i)}_{a}\right).
\end{equation*}
The symmetrization of the above equation is straightforward,
and one finds the characteristic polynomial
\begin{equation}
    \left(\sigma_{a_i}^{(i)}\right)^{k_{i+1}} + \sum_{r = 0}^{k_{i+1} - 1} \left(\sigma_{a_i}^{(i)}\right)^r g^{(i)}_{n-r,2d}\left(\sigma^{(i-1)}, \sigma^{(i+1)}, q_{i,2d}\right)
\: = \: 0,
\end{equation}
where
$g^{(i)}_{\ell,2d}$ is defined in equation~(\ref{COF}).
As a result,
the Vieta relations of the above equation will be equivalent to
equation~(\ref{Vieta}), and so again we recover the
quantum cohomology~(\ref{eq:genl-qh}).

\subsubsection{Comparison of quantum cohomology prediction to the literature}
\label{sect:compare-qh}

In this section we will argue that the non-equivariant specialization
of the result above matches the
non-equivariant quantum cohomology of \cite[theorem 0.2]{gk},
specialized to partial flag manifolds, as in \cite[example 2.2]{gk}.
The reader should note that the results in \cite{gk} are given in terms
of universal quotient bundles, which we convert to universal subbundles
here.

In our notation, the presentation of \cite[theorem 0.2]{gk} is
\begin{equation}
{\mathbb C}[\sigma^{(i)}_j, q_i ]^W / I_q,
\end{equation}
where $W$ is the Weyl group of $U(k_1) \times \cdots \times U(k_s)$ 
\begin{equation}
\omega \: = \: \prod_{i=1}^s \prod_{j<k}\left(  \sigma^{(i)}_j - \sigma^{(i)}_k \right),
\end{equation}
and the ideal is
\begin{equation}
I_q \: = \: \langle f \in {\mathbb C}[\sigma^{(i)}_j, q_i]^W \, | \,
\omega f \in J \rangle,
\end{equation}
for
\begin{equation}
J \: = \: \langle K_{ij}, i=1, \cdots s, j=1, \cdots, k_i \rangle,
\end{equation}
where
\begin{equation}
K_{ij} \: = \: 
\prod_{k=1}^{k_{i+1}} \left(  \sigma^{(i+1)}_k - \sigma^{(i)}_j \right)
\: + \:
(-)^{k_i} q_i \prod_{k=1}^{k_{i-1}} \left(  \sigma^{(i)}_j - 
\sigma^{(i-1)}_k\right),
\end{equation}
in conventions in which $k_0=0$, $k_{s+1} = n$, and \(\sigma^{(0)}=\sigma^{(s+1)} = 0\).

Next, we compute
\begin{eqnarray}
\sum_{j=1}^{k_i} K_{ij} (-)^j \left( \sigma^{(i)}_j \right)^a 
\prod_{\ell < k, \ell \neq j, k \neq j} \left(  \sigma^{(i)}_{\ell} - \sigma^{(i)}_k \right)
\: = \: A + B
\end{eqnarray}
for $0 \leq a \leq k_{i}-1$ and
where we define 
\begin{eqnarray}
A & = &
\sum_{j=1}^{k_i} (-)^j \left( \sigma^{(i)}_j \right)^a
\left(  \prod_{k=1}^{k_{i+1}} \left(  \sigma^{(i+1)}_k -
\sigma^{(i)}_j \right)
\right) \left(
\prod_{\ell < k, \ell \neq j, \ell \neq j} \left(
 \sigma^{(i)}_{\ell} - \sigma^{(i)}_k \right)
\right),
\\
B & = &
\sum_{j=1}^{k_i} (-)^j \left( \sigma^{(i)}_j \right)^a
\left( (-)^{k_i} q_i \prod_{k=1}^{k_{i-1}} \left( \sigma^{(i)}_j - 
\sigma^{(i-1)}_k\right)
\right)
 \left(
\prod_{\ell < k, \ell \neq j, \ell \neq j} \left(
 \sigma^{(i)}_{\ell} - \sigma^{(i)}_k \right)
\right).
\nonumber
\end{eqnarray}

Focusing momentarily on $A$,
one can show that
\begin{eqnarray}
A & = &
\sum_{j=1}^{k_i} 
\left[ 
\sum_{m=0}^{k_{i+1}} (-)^{m+j} e_{k_{i+1}-m}\left( \sigma^{(i+1)} \right)
\left( \sigma^{(i)}_j \right)^{a+m} \right]
\prod_{\ell<k, \ell \neq j, k \neq j}
\left( \sigma^{(i)}_{\ell} - \sigma^{(i)}_k \right),
\nonumber \\
& = &
\sum_{m=0}^{k_{i+1}} e_{k_{i+1}-m}\left( \sigma^{(i+1)} \right)
\sum_{j=1}^{k_i} (-)^{m+j}\left( \sigma_j^{(i)} \right)^{a+m}
\prod_{\ell < k, \ell \neq j, k \neq j} \left(  \sigma^{(i)}_{\ell}
- \sigma^{(i)}_k \right).
\end{eqnarray}

Next, we use the Jacobi-Trudi formula
\begin{equation}
s_{\lambda} \prod_{i<j}\left( x_i - x_j \right)
\: = \:
\det\left[\begin{array}{ccc}
x_1^{r-1+\lambda_1} & \cdots & x_r^{r-1+\lambda_1} \\
\vdots & & \vdots \\
x_1^{1+\lambda_r} & \cdots & x_r^{1+\lambda_r} \\
x_1^{\lambda_r} & \cdots & x_r^{\lambda_r} 
\end{array}
\right]
\end{equation}
for the Schur polynomial $s_{\lambda}(x_1, \cdots, x_r)$ of a 
Young tableau $\lambda$.
This implies
\begin{eqnarray}
\frac{
\sum_{j=1}^{k_i} (-)^{m+j}\left( \sigma_j^{(i)} \right)^{a+m}
 \prod_{\ell < k, \ell \neq j, k \neq j} \left( \sigma^{(i)}_{\ell}
 - \sigma^{(i)}_k \right)
}{
  \prod_{\ell < k} \left(  \sigma^{(i)}_{\ell} - \sigma^{(i)}_k \right)
}
& = &
(-)^{m-1} h_{a-k_i+m+1}\left( \sigma^{(i)} \right)
\nonumber
\end{eqnarray}
by using the Jacobi-Trudi formula for $\lambda = (a-k_i+m+1, 0, \cdots, 0)$,
for which $s_{\lambda} = h_{a-k_i+m+1}$.

Putting this together, we have, for $1\leq i \leq s$,
\begin{eqnarray}
\frac{A}{
\prod_{\ell < k} \left( \sigma^{(i)}_{\ell} - \sigma^{(i)}_k \right)
}
& = &
\sum_{m=0}^{k_{i+1}}(-)^{m-1} e_{k_{i+1}-m}\left( \sigma^{(i+1)} \right)
 h_{a-k_i+m+1}\left( \sigma^{(i)} \right).
\end{eqnarray}
Proceeding similarly, one can show
\begin{eqnarray}
\frac{B}{
\prod_{\ell < k} \left(  \sigma^{(i)}_{\ell} - \sigma^{(i)}_k \right)
}
& = &
q_i \sum_{m=0}^{k_{i-1}}(-)^{k_i+k_{i-1}-m-1} e_{k_{i-1}-m}\left( \sigma^{(i-1)} \right)
h_{m+a+1-k_i}\left( \sigma^{(i)} \right)
\end{eqnarray}
which is valid for all $1 \leq i \leq s$.

Let \(\alpha=a+k_{i+1}-k_i+1\), we have the following relations
\begin{multline}\label{eq:qh:Wei-Elana-rel}
    F_\alpha^{(i)}=-\sum_{m=0}^{k_{i+1}}(-)^{m}e_{k_{i+1}-m}(\sigma^{(i+1)})h_{m+\alpha-k_{i+1}}(\sigma^{(i)})\\-q_i\sum_{m=0}^{k_{i-1}}(-)^{k_i+k_{i-1}-m}e_{k_{i-1}-m}(\sigma^{(i-1)})h_{m+\alpha-k_{i+1}}(\sigma^{(i)})
\end{multline}
 for \(1\leq i\leq s\) and \(\alpha=k_{i+1}-k_i+1,\dots, k_{i+1}\).

Next, we compare to the quantum cohomology presentation we derived from
our predicted $\lambda_y$ relations in quantum K theory in 
section~\ref{sect:qh:exp-lambda},
for which the relations are
\begin{equation}
c({\mathcal S}_i) \star c({\mathcal S}_{i+1}/{\mathcal S}_i)
\: = \:
c( {\mathcal S}_{i+1}) + (-)^{k_i-k_{i-1}} q_i c({\mathcal S}_{i-1}),
\end{equation}
which in terms of symmetric polynomials in the $\sigma$'s can be written
\begin{equation} \label{eq:qh:noneq}
\sum_{r=0}^{\ell} e_r\left( \sigma^{(i)} \right)
e_{\ell-r}\left( \hat{\sigma}^{(i)} \right)
 \: = \:
e_{\ell}\left( \sigma^{(i+1)} \right)  \: + \:
 (-)^{k_i-k_{i-1}} q_i
e_{\ell-k_{i+1}+k_{i-1}}\left( \sigma^{(i-1)} \right)
\end{equation}
for $\ell = 0, 1, \dots, k_{i+1}$.

To that end, we multiply equation \eqref{eq:qh:noneq} by $y^\ell$ and sum over $\ell$ from $0$ to $k_{i+1}$, and obtain
\begin{multline}
    \left(\sum_{r=0}^{k_i} y^r e_r\left(\sigma^{(i)}\right) \right)\left(\sum_{t=0}^{k_{i+1}-k_i} y^t e_t\left(\hat{\sigma}^{(i)}\right) \right) \\= \left(\sum_{\ell=0}^{k_{i+1}} y^\ell e_\ell\left(\sigma^{(i+1)}\right)\right) + (-)^{k_i - k_{i-1}} q_i \left(\sum_{\ell = 0}^{k_{i+1}} y^\ell e_{\ell - k_{i+1} + k_{i-1} } \left(\sigma^{(i-1)}\right)\right)
\end{multline}
hence,
\begin{multline}
    \left(\sum_{\ell=0}^{k_{i+1}-k_i} y^\ell e_\ell\left(\hat{\sigma}^{(i)}\right) \right) = \left(\sum_{t=0}^{k_{i+1}} y^t e_t\left(\sigma^{(i+1)}\right)\right) \left(\sum_{r=0}^\infty (-)^r y^r h_r\left(\sigma^{(i)}\right)\right) \\
    + (-)^{k_i - k_{i-1}} q_i \left(\sum_{t = 0}^{k_{i+1}} y^t e_{t - k_{i+1} + k_{i-1} } \left(\sigma^{(i-1)}\right)\right) \left(\sum_{r=0}^\infty (-)^r y^r h_r\left(\sigma^{(i)}\right)\right)
\end{multline}
Now consider $y^\alpha$ terms for $k_{i+1} - k_i < \alpha \le k_{i+1}$, we have
\begin{multline}
    0 = \sum_{t=0}^{k_{i+1}} (-)^{\alpha - t} e_t\left(\sigma^{(i+1)}\right) h_{\alpha - t}\left(\sigma^{(i)}\right)\\
     + (-)^{k_i - k_{i-1}} q_i \sum_{t=0} (-)^{\alpha - t} e_{ t - k_{i+1} + k_{i-1} }\left(\sigma^{(i-1)}\right) h_{\alpha - t}\left(\sigma^{(i)}\right).
\end{multline}
After making change of summation variables and getting rid of a total sign $(-)^{\alpha - k_{i+1}}$, we obtain
\begin{multline}
    0 = \sum_{m=0}^{k_{i+1}} (-)^m e_{k_{i+1} - m}\left(\sigma^{(i+1)}\right) h_{m + \alpha - k_{i+1}} \left(\sigma^{(i)}\right)\\
    + (-)^{k_i + k_{i-1}} q_i \sum_{m=0}^{k_{i+1}} (-)^m e_{k_{i-1}-m}\left(\sigma^{(i-1)}\right) h_{m + \alpha - k_{i+1}}\left(\sigma^{(i)}\right)
\end{multline}
which agrees with equation \eqref{eq:qh:Wei-Elana-rel}.

This demonstrates that the non-equivariant specialization of
our prediction~(\ref{eq:genl-qh}) for quantum cohomology rings of partial
flag manifolds, holds.  In principle, to prove this rigorously, we would also
need to demonstrate that no additional relations are needed.
This follows from the graded Nakayama lemma,
as in \cite{mihalcea:giambelli}, see
Lemma 4.1 and Thm.~4.2; in turn, it generalizes to the equivariant version 
the statement from \cite[Prop. 11]{fulton.pandharipande}. 
More precisely, one needs to show that if one takes the classical 
specialization $q_i =0$ of the given relations, then they form a complete set of relations.
The latter fact follows from known presentations of partial flag manifolds when realized as 
towers of Grassmann bundles, see e.g.~\cite[Ex. 14.6.6]{fulton:IT}. 
We leave the details
to the reader.

\subsection{Duality}
\label{sect:fl:dual}

In this section we will apply 
the duality of flag manifolds relating
$F(k_1, \cdots, k_s, N)$ to $F(N-k_s, N-k_{s-1}, \cdots, N-k_1, N)$,
to write the $\lambda_y$ class relations in terms of universal quotient
bundles rather than universal subbundles.

Define ${\mathcal Q}_i = {\mathbb C}^N/S_i$, 
then we have the short exact sequences
\begin{equation} \label{eq:fl:seq1}
0 \: \longrightarrow \: {\mathcal S}_i \: \longrightarrow \: 
{\mathcal S}_{i+1} \: \longrightarrow
\: {\mathcal S}_{i+1}/{\mathcal S}_i \: \longrightarrow \: 0,
\end{equation}
\begin{equation}  \label{eq:fl:seq2}
0 \: \longrightarrow \: {\mathcal S}_{i+1}/{\mathcal S}_i 
\: \longrightarrow \: {\mathcal Q}_i
\: \longrightarrow \: {\mathcal Q}_{i+1} \: \longrightarrow \: 0.
\end{equation}
Using ${}^{\prime}$ to denote bundles
on the dual flag manifold, it is a classical result that
${\mathcal S}_i$ on the flag manifold $F(k_1, \cdots, k_s, N)$
is isomorphic to ${\mathcal Q}_{s+1-i}^{\prime *}$ on the dual flag manifold
$F(N-k_s, \cdots, N-k_1,N)$.  
Similarly, ${\mathcal S}_{i+1}/{\mathcal S}_i$ on the flag manifold
$F(k_1,\cdots, k_s,N)$ is isomorphic to
\begin{equation}
{\mathcal Q}_{s-i}^{\prime *} / {\mathcal Q}_{s+1-i}^{\prime *} \:
\cong \: \left( {\mathcal S}_{s+1-i}^{\prime} / {\mathcal S}_{s-1}^{\prime}
\right)^*
\end{equation}
on the dual flag manifold, where in the isomorphism we have used
the dual of the sequence~(\ref{eq:fl:seq2}).
This duality is also reflected in the two
short exact sequences above, as the sequence~(\ref{eq:fl:seq1}) maps to
\begin{equation}
0 \: \longrightarrow \: {\mathcal Q}_{s+1-i}^{\prime *}
\: \longrightarrow \: {\mathcal Q}_{s-i}^{\prime *} \: \longrightarrow \:
{\mathcal Q}_{s-i}^{\prime *}/{\mathcal Q}_{s+1-i}^{\prime *} \cong 
\left( {\mathcal S}^{\prime}_{s+1-i} / {\mathcal S}^{\prime}_{s-i} \right)^*
\: \longrightarrow \: 0,
\end{equation}
which is just the dual of sequence~(\ref{eq:fl:seq2}).

Now that we have seen how the duality relates the classical $\lambda_y$
relations, it remains to describe the quantum version.

Given the $\lambda_y$ relations in the form~(\ref{eq:qk:flag:reln:lambda}),
namely
\begin{equation} \nonumber
\lambda_y({\cal S}_i) \star \lambda_y({\cal S}_{i+1}/{\cal S}_i) \: = \:
\lambda_y({\cal S}_{i+1}) \: - \:
y ^{k_{i+1} - k_{i}} \frac{q_i}{1-q_i} \det ({\cal S}_{i+1}/{\cal S}_i) \star
\left( \lambda_y({\cal S}_{i}) - \lambda_y({\cal S}_{i-1}) \right),
\end{equation}
if we use primes ($\prime$) to denote bundles on the dual flag manifold,
we have immediately
\begin{eqnarray} \nonumber
\lefteqn{
\lambda_{y'}({\cal S}^{\prime}_i) \star \lambda_{y'}({\cal S}^{\prime}_{i+1}/{\cal S}^{\prime}_i)
}  \\
& = &
\lambda_{y'}({\cal S}^{\prime}_{i+1}) \: - \:
(y')^{(N-k_{s-i}) - (N-k_{s+1-i})} \frac{q'_i}{1-q'_i} \det ({\cal S}^{\prime}_{i+1}/{\cal S}^{\prime}_i) \star
\left( \lambda_{y'}({\cal S}^{\prime}_{i}) - \lambda_{y'}({\cal S}^{\prime}_{i-1}) \right).
\nonumber
\end{eqnarray}
Interpreting ${\cal S}^{\prime}_i = {\mathcal Q}_{s+1-i}^*$ as above,
and writing $j = s+1-i$, we can rewrite this in the form
\begin{eqnarray}
\lefteqn{
\lambda_{y'}({\mathcal Q}^*_j) \star
\lambda_{y'} \left( ( {\cal S}_j / {\cal S}_{j-1} )^* \right)
}  \\
& = &
\lambda_{y'}( {\mathcal Q}^*_{j-1} ) \: - \:
(y')^{k_j - k_{j-1}} \frac{q'_{s+1-j}}{1-q'_{s+1-j}}
\det\left( ( {\cal S}_j / {\cal S}_{j-1} )^* \right)
\star
\left( 
\lambda_{y'} ( {\mathcal Q}_j^* ) -
\lambda_{y'} ( {\mathcal Q}_{j+1}^* )
\right) 
\nonumber
\end{eqnarray}
on the original flag manifold.
We will not use this relation in this paper, but include it for
completeness.

\section{Examples}
\label{sect:exs}

\subsection{Incidence varieties, meaning, flag manifolds $F(1, n-1, n)$}
\label{sect:exs:inc}

\subsubsection{Overview of physics}

Now let us specialize to incidence varieties,
which are flag manifolds $F(1, n-1, n)$.
As described earlier in section~\ref{sect:flag:physics},
these are realized physically by
a $U(1) \times U(n-1)$ gauge theory with one bifundamental 
(charge $+1$ under $U(1)$ and antifundamental under $U(n-1)$),
and $n$ fundamentals of $U(n-1)$.

Generically on the Coulomb branch, the $U(n-1)$ gauge symmetry is
broken to a
$U(1)^{n-1}$ subgroup, 
with $(n-1)(n-2)$ W-bosons.
Specializing equation~(\ref{eq:W:full}), the superpotential in this
case is
\begin{eqnarray}
\mathcal{W} & = &
\frac{1}{2} (n-2) \sum_{a=1}^{n-1} \left( \ln X^{(2)}_a \right)^2
\: - \:
\sum_{i=1}^2 \sum_{1 \le a_i < b_i \le k_i} \left(\ln X^{(i)}_{a_i}\right) 
\left(\ln X^{(i)}_{b_i}\right)
\nonumber \\
& &
\: + \:
\left( \ln q_1 \right) \left( \ln X^{(1)} \right)
\: + \:
\left( \ln \left( (-)^{n-2} q_2 \right) \right) \sum_{a=1}^{n-1}
\left( \ln X^{(2)}_a \right)
\nonumber \\
& & 
\: + \: 
\sum_{a=1}^{n-1} {\rm Li}_2 \left( X^{(1)} / X^{(2)}_a \right)
\: + \:
\sum_{a=1}^{n-1} \sum_{i=1}^n {\rm Li}_2 \left( X^{(2)}_a / T_i \right).
\end{eqnarray}

The Coulomb branch equations~(\ref{eqn:QK-rel}), derived from derivatives
of the superpotential $\mathcal{W}$, specialize to
\begin{equation}
q_1  X^{(1)}  \: = \:
 X^{(1)}  \prod_{b=1}^{n-1} \left(
1 - \frac{ X^{(1)} }{ X^{(2)}_b } \right),
\end{equation}
\begin{equation}
(-)^{n-2} q_2 \left( X^{(2)}_a \right)^{n-1}
 \left( 1 - \frac{ X^{(1)} }{ X^{(2)}_a } 
\right)
\: = \:
\left( \prod_{b_2=1}^{n-1} X^{(2)}_{b_2} \right)
\prod_{b_3 = 1}^n \left( 1 -  \frac{ X^{(2)}_a }{ T_{b_3} } \right),
\end{equation}
for $a= 1, \cdots, n-1$,
which after cleaning up the algebra can be slightly rewritten
\begin{equation}
q_1 \: = \:  \prod_{b=1}^{n-1} \left(
1 - \frac{ X^{(1)} }{ X^{(2)}_b } \right),
\end{equation}
\begin{equation} 
(-)^{n-2} q_2 \left( X^{(2)}_a \right)^{n-2}
\left( X^{(2)}_a - X^{(1)} \right)
\: = \:
\left( \prod_{b_2=1}^{n-1} X^{(2)}_{b_2} \right)
\prod_{b_3 = 1}^n \left( 1 - \frac{ X^{(2)}_a }{  T_{b_3} } \right).
\end{equation}

After symmetrizing, these equations become 
(specializing~(\ref{eq:keyreln})):
\begin{equation}  \label{eq:inc:final:1}
\sum_{r=0}^{n-2} e_{\ell-r}\left( X^{(1)} \right) e_r\left( \overline{X}^{(1)}
\right) \: = \:
e_{\ell}\left( X^{(2)} \right) \: + \: 
\left\{ \begin{array}{cl}
q_1 e_{n-2}\left( \overline{X}^{(1)} \right) & \ell = n-2, 
\\
0 & {\rm else},
\end{array} \right.
\end{equation}
for $\ell = 0, \cdots, n-1$,
and
\begin{equation}  \label{eq:inc:final:2}
\sum_{r=0}^1 e_{\ell-r}\left( X^{(2)} \right) e_r\left( \overline{X}^{(2)}
\right) \: = \:
e_{\ell}(T) \: + \:
q_2 \, e_1\left( \overline{X}^{(2)} \right) e_{\ell-1}\left( X^{(1)} \right),
\end{equation}
for $\ell = 0, \cdots, n$.

Next, we turn to $\lambda_y$ class presentations.
We interpret $e_{\ell}( X^{(i)} )$ as $\wedge^{\ell} {\cal S}_i$,
and 
following the dictionary~(\ref{eq:dictionary}), we interpret
\begin{eqnarray}
e_{\ell}\left( \overline{X}^{(1)} \right)
& \leftrightarrow &
\left\{ \begin{array}{cl}
\wedge^{\ell}\left( {\cal S}_2/{\cal S}_1 \right) & \ell < n-2, \\
(1-q_1)^{-1} \det({\cal S}_2/{\cal S}_1) & \ell = n-2,
\end{array} \right.
\\
e_{\ell}\left( \overline{X}^{(2)} \right)
& \leftrightarrow &
\left\{ \begin{array}{cl}
1 & \ell = 0, \\
(1-q_2)^{-1} \, {\mathbb C}^n/{\cal S}_2 & \ell = 1.
\end{array} \right.
\end{eqnarray}
Then, equations~(\ref{eq:inc:final:1}), (\ref{eq:inc:final:2})
are interpreted as
\begin{eqnarray}
\lefteqn{
\sum_{r=0}^{n-3} \wedge^{\ell-r} {\cal S}_1 \, \star \, \wedge^{r} ({\cal S}_2/{\cal S}_1)
\: + \:
\frac{1}{1-q_1} \wedge^{\ell-(n-2)} {\cal S}_1 \, \star \, \det( {\cal S}_2/{\cal S}_1 )
} \nonumber \\
& \hspace*{1in} = &
\wedge^{\ell} {\cal S}_2 \: + \:
\left\{ \begin{array}{cl}
q_1 (1-q_1)^{-1} \det({\cal S}_2/{\cal S}_1) & \ell = n-2, \\
0 & {\rm else},
\end{array} \right.
\end{eqnarray}
and
\begin{equation}
\wedge^{\ell} {\cal S}_2 + \frac{1}{1-q_2} \wedge^{\ell-1} {\cal S}_2 \, \star \, {\mathbb C}^n/{\cal S}_2
\: = \: 
\wedge^{\ell} {\mathbb C}^n \: + \: \frac{q_2}{1-q_2} {\mathbb C}^n/{\cal S}_2
\, \star \, \wedge^{\ell-1} {\cal S}_1,
\end{equation}
respectively.

It is straightforward to rewrite these equations as the coefficients
of $y^{\ell}$ in the relations
\begin{eqnarray}
\lambda_y({\cal S}_1) \star \lambda_y({\cal S}_2/{\cal S}_1) & = &
\lambda_y({\cal S}_{2}) \: - \:
y^{n-2} \frac{ q_1 }{1-q_1} \det ({\cal S}_2/{\cal S}_1) \star
\left( \lambda_y({\cal S}_1) - 1 \right),
\label{eq:inc:bundle:1}
\\
\lambda_y({\cal S}_2) \star \lambda_y({\mathbb C}^n/{\cal S}_2) & = &
\lambda_y({\mathbb C}^n) \: - \:
y \frac{ q_2 }{1-q_2} \det ({\mathbb C}^n/{\cal S}_2) \star
\left( \lambda_y({\cal S}_2) - \lambda_y({\cal S}_1) \right),
\label{eq:inc:bundle:2}
\end{eqnarray}
where ${\cal S}_1$ has rank $k_1 = 1$ and ${\cal S}_2$ has rank $k_2 = n-1$, so that,
for example,
\begin{equation}
\det\left( {\mathbb C}^n/{\cal S}_2 \right) \: = \: {\mathbb C}^n/{\cal S}_2
\end{equation}
classically.

These are the specializations of the $\lambda_y$ class
relation~(\ref{eq:qk:flag:reln:lambda}) of section~\ref{sect:lambda-pres}
for the $T$-equivariant
quantum K theory ring relations, namely
\begin{equation}
\lambda_y({\cal S}_i) \star \lambda_y({\cal S}_{i+1}/{\cal S}_i) \: = \:
\lambda_y({\cal S}_{i+1}) \: - \:
y ^{k_{i+1} - k_{i}} \frac{q_i}{1-q_i} \det ({\cal S}_{i+1}/{\cal S}_i)  \star
\left( \lambda_y({\cal S}_{i}) - \lambda_y({\cal S}_{i-1}) \right).
\end{equation}

These $\lambda_y$ class relations for incidence varieties have also
been proven rigorously, using independent methods
from \cite{weihong}.  The proof will appear in 
\cite{mathpaper}.

\subsubsection{$F(1,2,3)$}
\label{sect:ex:f123}

In this section we will give explicitly the $T$-equivariant
quantum K theory relations for the special case $F(1,2,3)$.

Specializing the earlier analysis for incidence varieties,
the superpotential for $F(1,2,3)$ is
\begin{eqnarray}
\mathcal{W} & = &
\frac{1}{2}\sum_{a=1}^{2} \left( \ln X^{(2)}_a \right)^2
\: - \:
\left(\ln X^{(2)}_{1}\right) 
\left(\ln X^{(2)}_{2}\right)
\nonumber \\
& &
\: + \:
\left( \ln q_1 \right) \left( \ln X^{(1)} \right)
\: + \:
\left( \ln \left( (-) q_2 \right) \right) \sum_{a=1}^{2}
\left( \ln X^{(2)}_a \right)
\nonumber \\
& & 
\: + \: 
\sum_{a=1}^{2} {\rm Li}_2 \left( X^{(1)} / X^{(2)}_a \right)
\: + \:
\sum_{a=1}^{2} \sum_{i=1}^3 {\rm Li}_2 \left( X^{(2)}_a / T_i \right).
\end{eqnarray}
The Coulomb branch equations~(\ref{eqn:QK-rel}), derived from derivatives
of the superpotential $\mathcal{W}$, specialize to
\begin{equation}
q_1 \: = \:  \prod_{b=1}^{2} \left(
1 - \frac{ X^{(1)} }{ X^{(2)}_b } \right),
\end{equation}
\begin{equation} 
(-) q_2 \left( X^{(2)}_a \right)
\left( X^{(2)}_a - X^{(1)} \right)
\: = \:
\left( \prod_{b_2=1}^{2} X^{(2)}_{b_2} \right)
\prod_{b_3 = 1}^3 \left( 1 - \frac{ X^{(2)}_a }{  T_{b_3} } \right).
\end{equation}
After symmetrizing, these equations become
(specializing~(\ref{eq:keyreln})):
\begin{equation}  
\sum_{r=0}^{1} e_{\ell-r}\left( X^{(1)} \right) e_r\left( \overline{X}^{(1)}
\right) \: = \:
e_{\ell}\left( X^{(2)} \right) \: + \: 
\left\{ \begin{array}{cl}
q_1 e_{1}\left( \overline{X}^{(1)} \right) & \ell = 1, 
\\
0 & {\rm else},
\end{array} \right.
\end{equation}
for $\ell = 0, \cdots, 2$,
and
\begin{equation} 
\sum_{r=0}^1 e_{\ell-r}\left( X^{(2)} \right) e_r\left( \overline{X}^{(2)}
\right) \: = \:
e_{\ell}(T) \: + \:
q_2 \, e_1\left( \overline{X}^{(2)} \right) e_{\ell-1}\left( X^{(1)} \right),
\end{equation}
for $\ell = 0, \cdots, 3$.
Following the dictionary~(\ref{eq:dictionary}), we interpret
\begin{eqnarray}
e_{\ell}\left( \overline{X}^{(1)} \right)
& \leftrightarrow &
\left\{ \begin{array}{cl}
\wedge^{\ell}\left( {\cal S}_2/{\cal S}_1 \right) & \ell < n-2, \\
(1-q_1)^{-1} \det({\cal S}_2/{\cal S}_1) & \ell = n-2,
\end{array} \right.
\\
e_{\ell}\left( \overline{X}^{(2)} \right)
& \leftrightarrow &
\left\{ \begin{array}{cl}
1 & \ell = 0, \\
(1-q_2)^{-1} \, {\mathbb C}^n/{\cal S}_2 & \ell = 1.
\end{array} \right.
\end{eqnarray}
This leads to the $\lambda_y$ class relations, which, 
specializing~(\ref{eq:inc:bundle:1}), (\ref{eq:inc:bundle:2}),
take the form
\begin{eqnarray}
\lambda_y({\cal S}_1) \star \lambda_y({\cal S}_2/{\cal S}_1) & = &
\lambda_y({\cal S}_{2}) \: - \:
y \frac{ q_1 }{1-q_1} \det ({\cal S}_2/{\cal S}_1)  \star
\left( \lambda_y({\cal S}_1) - 1 \right),
\\
\lambda_y({\cal S}_2) \star \lambda_y({\mathbb C}^3/{\cal S}_2) & = &
\lambda_y({\mathbb C}^3) \: - \:
y \frac{ q_2 }{1-q_2} \det ({\mathbb C}^3/{\cal S}_2) \star
\left( \lambda_y({\cal S}_2) - \lambda_y({\cal S}_1) \right),
\end{eqnarray}
where ${\cal S}_1$ has rank $1$ and ${\cal S}_2$ has rank $2$.

We check these relations against known calculations in
the
quantum K ring utilizing Schubert classes. 
To reduce from the amount of notation, we 
will work in the non-equivariant situation. 

Fix the standard basis $e_1, e_2, e_3$ of $\mathbb{C}^3$. 
For a permutation $w \in S_3$ let $\ell(w)$ be the number of inversions of $w$. 
Denote by $X^w$ to be the Schubert variety of complex codimension $\ell(w)$
and which contains the torus fixed point 
$\langle e_{w(1)} \rangle \subset \langle e_{w(1)} , e_{w(2)}\rangle \subset \mathbb{C}^3$.
(See the companion paper \cite{mathpaper} for more details.) 
Denote by $\cO^w$ the class
in K theory given by the structure sheaf of $X^w$.
Denote by $s_1 = (12) $ and $s_2= (23)$
the simple reflections in $S_3$. To further simplify notation, we will denote by $\cO^1, 
\cO^{12}$ etc the Schubert classes associated to $s_1, s_1 s_2$ etc. 
With this notation
\begin{equation}\label{E:schur-to-schub} 
\begin{split} \cS_1& = 1 - \cO^1 \/;  \\
\cS_2/\cS_1 &= 1 + \cO^1 - \cO^2 - \cO^{1,2}\/; \\
\cS_2&=2 - \cO^2-\cO^{1,2} \/; \\
\wedge^2 \cS_2 & = 1- \cO^2\/; \\
\mathbb{C}^3/\cS_2 & = 1 + \cO^2+\cO^{1,2}  
\/.\end{split}
\end{equation}

The relevant multiplications in $\mathrm{QK}(\Fl(3))$ are given by 
(see e.g.~\cite[Theorem 5.1]{weihong}, \cite{Iritani:2013qka})
\[ \begin{split} \cO^1 \circ \cO^1 & = \cO^{2,1} -q_1 \cO^2 + q_1 \/; \\ 
\cO^1 \circ \cO^2 & = \cO^{1,2} + \cO^{2,1} - \cO^{1,2,1} \/; \\
\cO^1 \circ \cO^{1,2} & = \cO^{1,2,1} \/; \\
\cO^2 \circ \cO^2 & =  \cO^{1,2} -q_2 \cO^1 + q_2 \/; \\
\cO^2 \circ \cO^{1,2} & = q_2 \cO^{1} \/;\\
\cO^{1,2} \circ \cO^{1,2} & = q_2 \cO^{2,1} \/. \\
\end{split}
\]
It is straightforward to check that, with the dictionary above,
these products are equivalent to (the non-equivariant versions of)
the $\lambda_y$ class relations.

Let us now specialize the quantum cohomology relations~(\ref{eq:genl-qh})
to this case, and compare to rigorous mathematics results.  The $T$-equivariant
quantum
cohomology ring $QH_T^*(\Fl(3))$ has the relations\footnote{
In the mathematics literature, one often finds $T$-equivariant cohomology for
$T$ a maximal torus of $SL(n)$.  
Here, by contrast, we give results for $T$-equivariant
cohomology for $T$ a maximal torus of $GL(n)$ -- 
specifically, $GL(3)$.  We have checked that
our results also correctly reproduce results for equivariant cohomology
with respect to a torus in $SL(3)$, which corresponds to the special
case that
\begin{equation}
t_1 + t_2 + t_3 \: = \: 0.
\end{equation}
}
\begin{eqnarray} \label{eq:fl3:qh:1}
c^T({\cal S}_1) \star c^T\left( {\cal S}_2/{\cal S}_1 \right)
 & = & c^T({\cal S}_2) + (-) q_1,
\\
c^T( {\cal S}_2) \star c^T\left( {\mathbb C}^3/{\cal S}_2 \right) & = &
c^T( {\mathbb C}^3 ) + (-) q_2 \, c^T({\cal S}_1),
\label{eq:fl3:qh:2}
\end{eqnarray}
where $c^T$ denotes the $T$-equivariant total Chern class,
where the maximal torus $T$ consists of diagonal matrices in
$GL(3)$, and the $T$-module ${\mathbb C}^3$ has a weight space decomposition
${\mathbb C}^3 = {\mathbb C}_{t_1} \oplus {\mathbb C}_{t_2} \oplus
{\mathbb C}_{t_3}$, where $T$ acts on
${\mathbb C}_{t_i} \simeq {\mathbb C}$ with weight $t_i$.
For example,
\begin{equation}
c^T( {\mathbb C}^3) \: = \: (1 + t_1)(1 + t_2)(1 + t_3).
\end{equation}
We let $[X^w] \in H^{2 \ell(w)}_T(\Fl(3))$ denote the equivariant fundamental 
class indexed by $w$ (given by a reduced decomposition).
Then 
\[ 
\begin{split} 
c^T(\cS_1) &= 1 + t_1 - [X^1] \/, \\ 
c^T(\cS_2) &=  
(1+t_1)(1+t_2)-(1+t_1)[X^2] +[X^{1,2}]  \/, \\ 
c^T(\cS_2/\cS_1)& = (1 +t_2)  +[X^1]-[X^2] \/, \\
c^T(\mathbb{C}^3/\cS_2) & = 1 + t_3+ [X^2]  \/, 
\end{split}
\]
(The reader may recall that in taking the two-dimensional limit,
$T_i = \exp(L t_i)$.)

To check the relevant multiplications in $\mathrm{QH}^*_T(\mathrm{Fl}(3))$ one may use
the Chevalley formulae proved in \cite{mihalcea:eqqhom}. We obtain:
\[ 
\begin{split} 
[X^1] \star [X^1] & = [X^{2,1}]+(t_1 - t_2) [X^{1}]+q_1 \/, \\
[X^1] \star [X^2] &= [X^{2,1}] + [X^{1,2}] \/, \\
[X^2] \star [X^2] &= [X^{1,2}] + (t_2-t_3) [X^2] + q_2 \/, \\
[X^1] \star [X^{1,2}] & = (t_1-t_2) [X^{1,2}]+ [X^{1,2,1}] \/,\\
[X^2] \star [X^{1,2}] & = (t_1 - t_3) [X^{1,2}]+ q_2 [X^{1}] \/. 
\end{split}
\]
It can be shown that the products above correctly
reproduce the relations~(\ref{eq:fl3:qh:1}), (\ref{eq:fl3:qh:2}) arising
from physics.
Note that $\{ t_1- t_2, t_2-t_3 \}$ form a basis of the positive simple roots for the root system
associated to $\mathrm{GL}_3$.

As another consistency check,
it can also be shown that the non-equivariant version of the quantum cohomology
presentation above matches that in
\cite[theorem 0.2]{gk} for $F(1,2,3)$.

Now, let us compare to the presentation of the quantum K theory ring
of $F(1,2,3)$ in \cite[theorem 5.4]{Koroteev:2017nab} and \cite[theorem 3]{maeno.naito.sagaki:pres1}.
For $F(1,2,3)$, our presentation can be summarized as follows:
\begin{equation}\label{eqn:1}\cS_1+\cS_2/\cS_1=\cS_2\end{equation}
\begin{equation}\label{eqn:2}\cS_2+\C^3/\cS_2=\C^3\end{equation}
\begin{equation}\label{eqn:3}\cS_2/\cS_1\star\cS_1=(1-q_1)\wedge^2\cS_2\end{equation}
\begin{equation}\label{eqn:4}\C^3/\cS_2\star(\cS_2-q_2\cS_1)=(1-q_2)(\wedge^2\C^3-\wedge^2\cS_2)\end{equation}
\begin{equation}\label{eqn:5}\C^3/\cS_2\star\wedge^2\cS_2=(1-q_2)\wedge^3\C^3\end{equation}

To compare to \cite[theorem 5.4]{Koroteev:2017nab} and \cite[theorem 3]{maeno.naito.sagaki:pres1}, we define
\begin{equation*}\label{eqn:ps}
\mathfrak{p}_1= \cS_1 = (1-x_1)(1-q_1)\/,
\end{equation*}
\begin{equation*}
(1-q_1)\mathfrak{p}_2= \cS_2/\cS_1 = (1-q_2)(1-x_2) \/,\\
\end{equation*}
\begin{equation*}
(1-q_2) \mathfrak{p}_3=  \C^3/\cS_2 = 1- x_3 \/. 
\end{equation*}
Then by \ref{eqn:1} and \ref{eqn:2},
\begin{equation}
    \mathfrak{p}_1+\mathfrak{p}_2(1-q_1)+\mathfrak{p}_3(1-q_2)=\C^3;
\end{equation}
by \ref{eqn:1}, \ref{eqn:3} and \ref{eqn:4},
\begin{equation}
    \mathfrak{p}_1\star\mathfrak{p}_2+\mathfrak{p}_2\star\mathfrak{p}_3(1-q_1)+\mathfrak{p_3}\star\mathfrak{p}_1(1-q_2)=\wedge^2\C^3;
\end{equation}
and by \ref{eqn:3} and \ref{eqn:5},
\begin{equation}
    \mathfrak{p}_1\star\mathfrak{p}_2\star\mathfrak{p_3}=\wedge^3\C^3,
\end{equation}
recovering the presentation in \cite[Theorem 5.4]{Koroteev:2017nab}, 
where \(z^\#_i=q_i\) for \(i=1,2\), and \(e_r(\mathfrak{a}_1,\dots,\mathfrak{a}_3)=\wedge^r\C^3\) for \(r=1,2,3\).
Similarly,
\begin{equation*} 
(1-x_1)(1-q_1) + (1-x_2)(1-q_2) + (1-x_3) = \C^3 \/,
\end{equation*}
\begin{equation*} (1-x_1)(1-x_2)(1-q_2) + (1-x_1)(1-x_3)(1-q_1) + (1-x_2)(1-x_3) = \wedge^2 \C^3 \/,
\end{equation*}
\begin{equation*} (1-x_1)(1-x_2)(1-x_3) = \wedge^3 \C^3  \/. 
 \end{equation*}
recovering the presentation in \cite[theorem 3]{maeno.naito.sagaki:pres1}.
\subsection{Full flag manifolds}

\subsubsection{Overview}

Consider the full flag manifold $F(1,2,3,\cdots,n)$,
(meaning that $k_i = i$ for each $i$, and all steps appear,)
and let $S_k$ be the $k$th universal subbundle, which has rank $k$.
As described earlier in section~\ref{sect:flag:physics},
these are realized physically by a $U(1) \times U(2) \times \cdots
\times U(n-1)$ gauge theory with a set of bifundamentals.

Generically on the Coulomb branch, each $U(k)$ gauge symmetry factor is broken
to a $U(1)^k$ subgroup, with $k(k-1)$ W-bosons.
Specializing equation~(\ref{eq:W:full}), the superpotential in this
case is
\begin{eqnarray}
\mathcal{W} & = &
\frac{1}{2} \sum_{i=1}^{n-1} (i - 1) \sum_{a_i=1}^i \left( \ln X^{(i)}_{a_i}
\right)^2 \: - \:
\sum_{i=1}^{n-1} \sum_{1 \leq a_i < b_i \leq i} 
\left( \ln X^{(i)}_{a_i} \right)
\left( \ln X^{(i)}_{b_i} \right)
\nonumber \\
& &
\: + \:
\sum_{i=1}^{n-1} \left( \ln\left( (-)^{i-1} q_i \right) \right)
\sum_{a_i=1}^i \left( \ln X^{(i)}_{a_i} \right)
\nonumber \\
& & 
\: + \:
\sum_{i=1}^{n-1} \sum_{a_i=1}^{i} \sum_{a_{i+1}=1}^{i+1}
{\rm Li}_2 \left( X^{(i)}_{a_i} / X^{(i+1)}_{a_{i+1}} \right),
\end{eqnarray}
where $k_0 = 0$, $k_n = n$, $X^{(n)}_{a_n} = T_{a_n}$,
and we have used the fact that there are $s = n-1$ gauge factors and
each rank $k_i = i$.

The Coulomb branch equations~(\ref{eqn:QK-rel}), derived from derivatives
of the superpotential $\mathcal{W}$, specialize to
\begin{equation}
(-)^{i-1} q_i \left( X^{(i)}_{a_i} \right)^i \prod_{b_{i-1}=1}^{i-1}
\left( 1 - \frac{ X^{(i-1)}_{b_{i-1}} }{ X^{(i)}_{a_i} } \right)
\: = \:
\left( \prod_{b_i=1}^i x^{(i)}_{b_i} \right)
\prod_{b_{i+1}=1}^{i+1} \left( 1 -
\frac{ X^{(i)}_{a_i} }{ X^{(i+1)}_{b_{i+1}} } \right),
\end{equation}
for $a_i = 1, \cdots, i$ and $i=1, \cdots, n-1$.

After symmetrizing, these equations become (specializing~(\ref{eq:keyreln}))
\begin{equation} \label{eq:full:final}
\sum_{r=0}^1 e_{\ell-r}\left( X^{(i)} \right)
e_r\left( \overline{X}^{(i)} \right)
\: = \: 
e_{\ell}\left( X^{(i+1)} \right) \: + \:
q_i \, e_1\left( \overline{X}^{(i)} \right)
e_{\ell-1}\left( X^{(i-1)} \right),
\end{equation}
in conventions in which $X^{(n)} = T$ and $X^{(0)} = 0$.

We interpret $e_{\ell}( X^{(i)} )$ as $\wedge^{\ell} {\cal S}_i$, 
identifying the components $X^{(i)}_a$ with Chern roots of ${\cal S}_i$,
and
following the dictionary~(\ref{eq:dictionary}), we interpret
\begin{equation}
e_{\ell}\left( \overline{X}^{(i)} \right)
\: \leftrightarrow \:
\left\{ \begin{array}{cl}
{\cal O} & \ell = 0,
\\
(1-q_i)^{-1} {\cal S}_{i+1}/{\cal S}_i & \ell = 1.
\end{array} \right.
\end{equation}
(Since ${\cal S}_{i+1}/{\cal S}_i$ has rank one, there are no higher exterior powers,
and ${\cal S}_{i+1}/{\cal S}_i = \det {\cal S}_{i+1}/{\cal S}_i$.)
Then, equation~(\ref{eq:full:final}) becomes
\begin{equation}
\wedge^{\ell} {\cal S}_i \: + \: \frac{1}{1-q_i} \wedge^{\ell-1} {\cal S}_i \star ({\cal S}_{i+1}/{\cal S}_i)
\: = \: 
\wedge^{\ell} {\cal S}_{i+1} \: + \:
\frac{ q_i }{1-q_i} ({\cal S}_{i+1}/{\cal S}_i) \, \star \,
 \wedge^{\ell-1} {\cal S}_{i-1}.
\end{equation}
which can be rearranged algebraically to become
\begin{equation}
\sum_{s=0}^1 \wedge^{\ell-s} {\cal S}_i \, \star \,
 \wedge^s ({\cal S}_{i+1}/{\cal S}_i) \: = \:
\wedge^{\ell} {\cal S}_{i+1} \: - \:
\frac{q_i}{1-q_i} \det ({\cal S}_{i+1}/{\cal S}_i)  \star \left(
\wedge^{\ell-1} {\cal S}_i - \wedge^{\ell-1} {\cal S}_{i-1} \right).
\end{equation}

Adding powers of $y$, this can be encoded in the $\lambda_y$ class
expression
\begin{equation}
\lambda_y({\cal S}_i) \star \lambda_y({\cal S}_{i+1}/{\cal S}_i) \: = \:
\lambda_y({\cal S}_{i+1}) \: - \: y \frac{q_i}{1-q_i} 
\det ({\cal S}_{i+1}/{\cal S}_i)  \star
 \left( \lambda_y({\cal S}_i) - \lambda_y({\cal S}_{i-1}) \right),
\end{equation}
which is precisely the specialization of the $\lambda_y$ class
relation~(\ref{eq:qk:flag:reln:lambda}).

A different presentation may also be helpful.
As discussed earlier in section~\ref{sect:lambda-pres}, 
the $\lambda_y$ class relations imply~(\ref{eqn:rel2}), (\ref{eqn:rel1}),
of which the relation~(\ref{eqn:rel1}) specializes in the present case to
\begin{equation} \label{eq:lambda:master}
\left( \wedge^k {\cal S}_k \right) \otimes \left( \wedge^{\ell} {\cal S}_{k+1} - 
\wedge^{\ell} {\cal S}_k \right) \: = \:
\left( \wedge^{\ell-1} {\cal S}_k - q_k \wedge^{\ell-1} {\cal S}_{k-1} \right)
\otimes \wedge^{k+1} {\cal S}_{k+1}.
\end{equation}
The relation~(\ref{eqn:rel2}) is redundant, as it corresponds to the special
case that $\ell = 1$.

\subsubsection{$F(1,2,3,4)$}
\label{sect:ex:f1234}

As we have already considered $F(1,2,3)$ in section~\ref{sect:ex:f123},
we turn here to the next full flag manifold, namely,
$F(1,2,3,4)$ with tautological subbundles
\begin{equation}
0 = \cS_0 \subset \cS_1 \subset \cS_2 \subset \cS_3 \subset \cS_4 = \mathbb{C}^4,
\end{equation}
of ranks $0, 1, 2, 3, 4$ respectively.

Specializing equation~(\ref{eq:W:full}), the superpotential in this
case is
\begin{eqnarray}
\mathcal{W} & = &
\frac{1}{2} \sum_{i=1}^{3} (i - 1) \sum_{a_i=1}^i \left( \ln X^{(i)}_{a_i}
\right)^2 \: - \:
\sum_{i=1}^{3} \sum_{1 \leq a_i < b_i \leq i} 
\left( \ln X^{(i)}_{a_i} \right)
\left( \ln X^{(i)}_{b_i} \right)
\nonumber \\
& &
\: + \:
\sum_{i=1}^{3} \left( \ln\left( (-)^{i-1} q_i \right) \right)
\sum_{a_i=1}^i \left( \ln X^{(i)}_{a_i} \right)
\nonumber \\
& & 
\: + \:
\sum_{i=1}^{3} \sum_{a_i=1}^{i} \sum_{a_{i+1}=1}^{i+1}
{\rm Li}_2 \left( X^{(i)}_{a_i} / X^{(i+1)}_{a_{i+1}} \right).
\end{eqnarray}
The Coulomb branch equations~(\ref{eqn:QK-rel}), derived from derivatives
of the superpotential $\mathcal{W}$, specialize to
\begin{equation}
(-)^{i-1} q_i \left( X^{(i)}_{a_i} \right)^i \prod_{b_{i-1}=1}^{i-1}
\left( 1 - \frac{ X^{(i-1)}_{b_{i-1}} }{ X^{(i)}_{a_i} } \right)
\: = \:
\left( \prod_{b_i=1}^i x^{(i)}_{b_i} \right)
\prod_{b_{i+1}=1}^{i+1} \left( 1 -
\frac{ X^{(i)}_{a_i} }{ X^{(i+1)}_{b_{i+1}} } \right),
\end{equation}
for $a_i = 1, \cdots, i$ and $i=1, \cdots, 3$.
After symmetrizing, these equations become (specializing~(\ref{eq:keyreln}))
\begin{equation} 
\sum_{r=0}^1 e_{\ell-r}\left( X^{(i)} \right)
e_r\left( \overline{X}^{(i)} \right)
\: = \: 
e_{\ell}\left( X^{(i+1)} \right) \: + \:
q_i \, e_1\left( \overline{X}^{(i)} \right)
e_{\ell-1}\left( X^{(i-1)} \right),
\end{equation}
and after interpreting as before, we are led to $\lambda_y$ class
relations (`quantum Whitney relations') given by
\begin{equation}
\lambda_y({\cal S}_i) \star \lambda_y({\cal S}_{i+1}/{\cal S}_i) \: = \:
\lambda_y({\cal S}_{i+1}) \: - \: y \frac{q_i}{1-q_i} 
\det ({\cal S}_{i+1}/{\cal S}_i)  \star
 \left( \lambda_y({\cal S}_i) - \lambda_y({\cal S}_{i-1}) \right),
\end{equation}
for $1 \le i \le 3$, i.e.,
\[ \begin{split} \lambda_y({\cal S}_1) \star \lambda_y({\cal S}_{2}/{\cal S}_1) &= 
\lambda_y({\cal S}_{2}) \: - \: y \frac{q_1}{1-q_1} 
\det ({\cal S}_{2}/{\cal S}_1)  \star
 \left( \lambda_y({\cal S}_1) - 1 \right) \/, \\
 \lambda_y({\cal S}_2) \star \lambda_y({\cal S}_{3}/{\cal S}_2) &= 
\lambda_y({\cal S}_{3}) \: - \: y \frac{q_2}{1-q_2} 
\det ({\cal S}_{3}/{\cal S}_2)  \star
 \left( \lambda_y({\cal S}_2) - \lambda_y(\cS_1) \right) \/, \\
\lambda_y({\cal S}_3) \star \lambda_y(\mathbb{C}^4/{\cal S}_3) &= 
\lambda_y(\mathbb{C}^4) \: - \: y \frac{q_3}{1-q_3} 
\det (\mathbb{C}^4/{\cal S}_3)  \star
 \left( \lambda_y({\cal S}_3) - \lambda_y(\cS_2) \right) \/.
 \end{split}
 \]
This is the specialization of the $\lambda_y$ class
relation~(\ref{eq:qk:flag:reln:lambda}).

Now, let us compare to existing\footnote{
Conjectural formulas for the Chevalley coefficients appeared in
\cite{lenart.postnikov:affine}, and were later proven in \cite{lns}.
} mathematics results, which are phrased
in terms of Schubert classes.
(As for $F(1,2,3)$, we will only compare to non-equivariant quantum K theory.)
The Schubert classes will be denoted by $\cO^w$ with $w \in S_4$, and the same
 definitions as those for $\Fl(3) = F(1,2,3)$ are utilized. 
For the convenience of the reader, we
 list the expansions of the Schubert classes of each exterior power of the bundles $\cS_i$.
\begin{equation} 
\begin{split} 
\cS_1& = 1 - \cO^1\/,  \\
\cS_2&=2 - \cO^2-\cO^{1,2} \/, \\
\wedge^2 \cS_2 & = 1- \cO^2\/, \\
\cS_3 & = 3- \cO^3 - \cO^{2, 3} - \cO^{1, 2, 3}\/,\\
\wedge^2 \cS_3 & = 3 - 2 \cO^3 - \cO^{2,3}\/, \\
\wedge^3 \cS_3 & = 1 - \cO^3 \/, \\ 
\cS_i/\cS_{i-1} & = \cS_i - \cS_{i-1} \/, \quad 1 \le i \le 4 \/.
\end{split}
\end{equation}
The relevant multiplications of Schubert classes are:
\[ \begin{split} \cO^{1} \circ \cO^{1} & = \cO^{2,1}-q_1\cO^{2}+q_1 \/, \\ 
\cO^1 \circ \cO^2 & = \cO^{1,2} + \cO^{2,1} - \cO^{1,2,1} \/, \\
\cO^1 \circ \cO^{3} & = \cO^{1,3} \/, \\
\cO^1 \circ \cO^{1,2} & = \cO^{1,2,1} \/, \\
\cO^1 \circ \cO^{2,3} & = \cO^{1,2,3}+ \cO^{2,3,1}-\cO^{1,2,3,1} \/, \\
\cO^1 \circ \cO^{1,2,3} & =\cO^{1,2,3,1} \/, \\
\cO^2 \circ \cO^2 & =  \cO^{1,2}+\cO^{3,2}-\cO^{3,1,2} -q_2 \cO^1 + q_2 \cO^{1,3}- q_2 \cO^{3}+ q_2 \/, \\
\cO^2 \circ \cO^3 & =\cO^{2,3}+ \cO^{3,2} - \cO^{2,3,2} \/,\\
\cO^2 \circ \cO^{1,2} & = \cO^{3,1,2} - q_2 \cO^{1,3}+q_2 \cO^{1} \/,\\
\cO^2 \circ \cO^{2,3} & = \cO^{1,2,3}+ \cO^{2,3,2}-\cO^{1,2,3,2} \/, \\
\cO^2 \circ \cO^{1,2,3} & = \cO^{1,2,3,2} \/, \\
\cO^3 \circ \cO^3 & = \cO^{2,3}+q_3-q_3\cO^{2} \/, \\
\cO^3 \circ \cO^{1,2} & =\cO^{3,1,2} + \cO^{1,2,3} - \cO^{1,2,3,2} \/, \\
\cO^3 \circ \cO^{2,3} & = \cO^{1,2,3} +q_3 \cO^2 - q_3 \cO^{1,2}\/, \\
\cO^3 \circ \cO^{1,2,3} & = q_3 \cO^{1,2} \/, \\
\cO^{1,2} \circ \cO^{1,2} & = \cO^{2,3,1,2} +q_2 \cO^{2,1}-q_2 \cO^{2,3,1} \/, \\
\cO^{1,2} \circ \cO^{2,3} & = \cO^{1,2,3,2}+\cO^{2,3,1,2}-\cO^{1,2,3,1,2} \/, \\
\cO^{1,2} \circ \cO^{1,2,3} & = \cO^{1,2,3,1,2}\/, \\
\cO^{2,3} \circ \cO^{2,3} & =q_3\cO^{3,2}+q_3\cO^{1,2}-q_3\cO^{3,1,2} \/, \\
\cO^{2,3} \circ \cO^{1,2,3} & =q_3 \cO^{3,1,2} \/, \\
\cO^{1,2,3} \circ \cO^{1,2,3} & =q_3 \cO^{2,3,1,2} \/.
\end{split}
\]
It is straightforward, albeit tedious, to check that the multiplications above
are consistent with the (nonequivariant version of the) 
$\lambda_y$ class relations for $T$-equivariant quantum K theory.

Let us now specialize the quantum cohomology relations~(\ref{eq:genl-qh})
to this case, and compare to rigorous mathematics results.  The 
$T$-equivariant quantum
cohomology ring $QH_T^*(\Fl(4))$ has the relations
\begin{eqnarray}
c^T({\cal S}_1) \star c^T\left( {\cal S}_2/{\cal S}_1\right)
& = &
c^T\left( {\cal S}_2 \right) \: + \: (-) q_1,
\label{eq:fl4:qh:1}\\
c^T({\cal S}_2) \star c^T\left( {\cal S}_3/{\cal S}_2 \right)
& = &
c^T\left( {\cal S}_3 \right) \: + \: (-) q_2 \, c^T\left( {\cal S}_1 \right),
\label{eq:fl4:qh:2}\\
c^T({\cal S}_3) \star c^T\left( {\mathbb C}^4/{\cal S}_3 \right)
& = &
c^T\left( {\mathbb C}^4 \right) \: + \: (-) q_3 \,
c^T\left( {\cal S}_2 \right),
\label{eq:fl4:qh:3}
\end{eqnarray}
where $c^T$ denotes the $T$-equivariant total Chern class.

Next, we compare to the literature.
In the same notation as section~\ref{sect:ex:f123}, 
we let $[X^w] \in H^{2 \ell(w)}_T(\Fl(3))$ denote the equivariant fundamental
class indexed by $w$ (given by a reduced decomposition). The maximal torus $T$
consists of diagonal matrices in $\mathrm{GL}_4$, and the $T$-module $\mathbb{C}^4$ has 
a weight space decomposition $\mathbb{C}^4 = \mathbb{C}_{t_1} \oplus  \mathbb{C}_{t_2} \oplus  
\mathbb{C}_{t_3} \oplus  \mathbb{C}_{t_4}$
where $T$ acts on $\mathbb{C}_{t_i} \simeq \mathbb{C}$ with weight $t_i$. 
Then
\[ 
\begin{split} 
c^T(\cS_1) &=
  (1 + t_1) - [X^1] \/, \\
c^T(\cS_2) &=  
[X^{1,2}] -(1+t_1) [X^2] + (1+t_1)(1+t_2) \/, \\
c^T(\cS_3) & = 
- [X^{1,2,3}]  -(1+t_1)(1+t_2) [X^{3}]+(1+t_1) [X^{2,3}] \\ &+
(1+t_1)(1+t_2)(1+t_3) \/, \\
c^T(\cS_2/\cS_1)& = 
(1 +t_2) + [X^1] - [X^2] \/, \\
c^T(\cS_3/\cS_2) & = 
(1 +t_3) + [X^2] - [X^3] \/, \\
c^T(\mathbb{C}^4/\cS_3) & = 
(1 +t_4) + [X^3] \/. \\
\end{split}
\]

To check the relevant multiplications in $\mathrm{QH}^*_T(\mathrm{Fl}(3))$ one may use
the Chevalley formulae proved in \cite{mihalcea:eqqhom}. We obtain
\[ 
\begin{split} 
[X^1] \star [X^1] & = [X^{2,1}]+(t_1 - t_2) [X^{1}]+q_1 \/, \\
[X^1] \star [X^2] &= [X^{2,1}] + [X^{1,2}] \/, \\
[X^1] \star [X^3] &= [X^{3,1}] \/, \\
[X^2] \star [X^2] &= [X^{1,2}]+[X^{3,2}]+(t_2-t_3)[X^2]+q_2 \/, \\
[X^2] \star [X^3] &= [X^{2,3}] + [X^{3,2}] \/, \\
[X^3] \star [X^3] & = [X^{2,3}]+(t_3 - t_4) [X^{3}]+q_3 \/, \\
[X^1] \star [X^{1,2}] & = (t_1 - t_2) [X^{1,2}]+ [X^{1,2,1}] \/,\\
[X^2] \star [X^{1,2}] & = (t_1-t_3) [X^{1,2}]+ q_2 [X^{1}]+[X^{3,1,2}] \/,\\
[X^3] \star [X^{1,2}] & = [X^{3,1,2}]+ [X^{1,2,3}] \/,\\
[X^3] \star [X^{2,3}] & =  (t_2-t_4) [X^{2,3}]+ q_3 [X^{2}]+ [X^{1,2,3}] \/,\\
[X^3] \star [X^{1,2,3}] & =q_3[X^{1,2}] + (t_1-t_4) [X^{1,2,3}] \/.
\end{split}
\]
It is straightforward to check that these products do indeed reproduce
the relations~(\ref{eq:fl4:qh:1}), (\ref{eq:fl4:qh:2}), 
and (\ref{eq:fl4:qh:3}).
Note that $\{ t_1- t_2, t_2-t_3, t_3-t_4 \}$ form a basis of the positive simple roots for the root system
associated to $\mathrm{GL}_4$.

\section{Conclusions}

In this paper we have used Coulomb branch methods in GLSMs to make
predictions for quantum K theory ring relations in partial flag
manifolds.

As remarked in section~\ref{sect:lambda-pres},
the form of our result for partial flag manifolds was suggestive
of an alternative expression in terms of a relative Grassmannian,
suggesting that there might exist a notion of ``vertical quantum K theory.''
We leave this for future work.

In discussions of duality between Grassmannians and flag manifolds and
their duals in sections~\ref{sect:gr:dual} and \ref{sect:fl:dual}, 
we found an elegant expression for certain dual bundles in terms
of elements of K theory which differ from honest bundles by factors of $q$.
This suggests that there might exist a ``quantum duality'' map,
a map ${\cal E} \mapsto {\cal E}^*$ for a `quantum' dual operation $*$.
We leave this for future work.

\section{Acknowledgements}

We would like to thank R.~Donagi, P.~Koroteev, 
and J.~Knapp for useful discussions.
W.G.~was partially supported by NSF grant PHY-1720321.
L.M.~ was partially supported by NSF grant DMS-2152294 and a Simons Collaboration 
grant; E.S.~was partially supported by NSF grant PHY-2014086. H.Z.~was supported by the China Postdoctoral Science Foundation with grant No.~2022M720509.


\begin{thebibliography}{199}

\addcontentsline{toc}{section}{References}

\bibitem{Bullimore:2014awa}
M.~Bullimore, H.~C.~Kim and P.~Koroteev,
``Defects and quantum Seiberg-Witten geometry,''
JHEP \textbf{05} (2015) 095,
{\tt arXiv:1412.6081 [hep-th]}.



\bibitem{Jockers:2018sfl}
H.~Jockers and P.~Mayr,
``A 3d gauge theory/quantum K-theory correspondence,''
Adv. Theor. Math. Phys. \textbf{24} (2020) 327-457,
{\tt arXiv:1808.02040 [hep-th]}.

\bibitem{Jockers:2019wjh}
H.~Jockers and P.~Mayr,
``Quantum K-theory of Calabi-Yau manifolds,''
JHEP \textbf{11} (2019) 011,
{\tt arXiv:1905.03548 [hep-th]}.

\bibitem{Jockers:2019lwe}
H.~Jockers, P.~Mayr, U.~Ninad and A.~Tabler,
``Wilson loop algebras and quantum K-theory for Grassmannians,''
JHEP \textbf{10} (2020) 036,
{\tt arXiv:1911.13286 [hep-th]}.


\bibitem{Gu:2020zpg}
W.~Gu, L.~Mihalcea, E.~Sharpe and H.~Zou,
``Quantum K theory of symplectic Grassmannians,''
J. Geom. Phys. \textbf{177} (2022) 104548,
{\tt arXiv:2008.04909 [hep-th]}.

\bibitem{Gu:2022yvj}
W.~Gu, L.~C.~Mihalcea, E.~Sharpe and H.~Zou,
``Quantum K theory of Grassmannians, Wilson line operators, and Schur bundles,''
{\tt arXiv:2208.01091 [math.AG]}.




\bibitem{Ueda:2019qhg}
K.~Ueda and Y.~Yoshida,
``3d $ \mathcal{N} $ = 2 Chern-Simons-matter theory, Bethe ansatz, and quantum $K$-theory of Grassmannians,''
JHEP \textbf{08} (2020) 157,
{\tt arXiv:1912.03792 [hep-th]}.


\bibitem{Morrison:1994fr}
D.~R.~Morrison and M.~R.~Plesser,
``Summing the instantons: Quantum cohomology and mirror symmetry in toric varieties,''
Nucl. Phys. B \textbf{440} (1995) 279-354,
{\tt arXiv:hep-th/9412236 [hep-th]}.

\bibitem{Koroteev:2017nab}
P.~Koroteev, P.~P.~Pushkar, A.~V.~Smirnov and A.~M.~Zeitlin,
``Quantum K-theory of quiver varieties and many-body systems,''
Selecta Math. New Ser. \textbf{27} (2021) 87,
{\tt arXiv:1705.10419 [math.AG]}.

\bibitem{Givental:1993nc}
A.~Givental and B.~s.~Kim,
``Quantum cohomology of flag manifolds and Toda lattices,''
Commun. Math. Phys. \textbf{168} (1995) 609-642,
{\tt arXiv:hep-th/9312096 [hep-th]}.

\bibitem{kim1} B. Kim,
``Quantum cohomology of flag manifolds $G/B$ and quantum Toda lattices,''
Ann. of Math. (2) {\bf 149} (1999) 129-148,
{\tt arXiv:alg-geom/9607001}.

\bibitem{Givental:2001clq}
A.~Givental and Y.~P.~Lee,
``Quantum K-theory on flag manifolds, finite-difference Toda lattices and quantum groups,''
Invent. Math. \textbf{151} (2003) 193-219,
{\tt arXiv:math/0108105 [math.AG]}.

\bibitem{Koroteev:2021lvp}
P.~Koroteev and A.~M.~Zeitlin,
``3d mirror symmetry for instanton moduli spaces,''
{\tt arXiv:2105.00588 [math.AG]}.



\bibitem{mns} T. Maeno, S. Naito, D. Sagaki,
``A presentation of the torus-equivariant quantum K-theory ring of flag
manifolds of type A, part II: quantum double Grothendieck polynomials,''
{\tt arXiv:2305.17685 [math.QA]}.

\bibitem{act} D. Anderson, L. Chen, H.-H. Tseng,
``On the quantum K-ring of the flag manifold,''
{\tt arXiv:1711.08414 [math.AG]}.


\bibitem{gkbethe} V. Gorbounov, C. Korff, ``Quantum integrability and generalised
quantum Schubert calculus,''
Adv. Math. {\bf 313} (2017) 282-356,
{\tt arXiv:1408.4718 [math.RT]}.



\bibitem{maeno.naito.sagaki:pres1}
T.~Maeno, S.~Naito, D.~Sagaki, ``A presentation of the torus equivariant
quantum K theory ring of flag manifolds of type A, Part I: the defining ideal",
{\tt arXiv:2302.09485 [math.QA]}.

\bibitem{lns} 
C.~Lenart, S.~Naito, D.~Sagaki,
``A general Chevalley formula for semi-infinite flag manifolds and
quantum K-theory,''  
{\tt arXiv:2010.06143 [math.CO]}.




\bibitem{mathpaper} W. Gu, L. Mihalcea, E. Sharpe, W. Xu, H. Zhang,
H. Zou, ``The quantum K Whitney relations for partial flag varieties,''
{\tt arXiv:2310.03826 [math.AG]}.

\bibitem{Guo:2018iyr} 
J.~Guo,
``Quantum sheaf cohomology and duality of flag manifolds,''
Commun. Math. Phys. \textbf{374} (2019)  661-688,
{\tt arXiv:1808.00716 [hep-th]}.

\bibitem{Bonelli:2013mma}
G.~Bonelli, A.~Sciarappa, A.~Tanzini and P.~Vasko,
``Vortex partition functions, wall crossing and equivariant Gromov-Witten invariants,''
Commun. Math. Phys. \textbf{333} (2015)  717-760,
{\tt arXiv:1307.5997 [hep-th]}.



\bibitem{Closset:2016arn}
C.~Closset and H.~Kim,
``Comments on twisted indices in 3d supersymmetric gauge theories,''
JHEP \textbf{08} (2016) 059,
{\tt arXiv:1605.06531 [hep-th]}.

\bibitem{Nekrasov:2009uh}
N.~A.~Nekrasov and S.~L.~Shatashvili,
``Supersymmetric vacua and Bethe ansatz,''
Nucl. Phys. B Proc. Suppl. \textbf{192-193} (2009) 91-112,
{\tt arXiv:0901.4744 [hep-th]}.







\bibitem{Hori:2011pd}
K.~Hori,
``Duality in two-dimensional (2,2) supersymmetric non-abelian gauge theories,''
JHEP \textbf{10} (2013) 121,
{\tt arXiv:1104.2853 [hep-th]}.

\bibitem{Gu:2018fpm}
W.~Gu and E.~Sharpe,
``A proposal for nonabelian mirrors,''
{\tt arXiv:1806.04678 [hep-th]}.



\bibitem{DiFrancesco:1997nk}
P.~Di Francesco, P.~Mathieu and D.~Senechal,
{\it Conformal field theory},
Springer-Verlag, New York, 1997.

\bibitem{hirzebruch} F. Hirzebruch,
{\it Topological methods in algebraic geometry}, third edition,
Springer-Verlag, New York, 1966. 



\bibitem{Donagi:2007hi}
R.~Donagi and E.~Sharpe,
``GLSM's for partial flag manifolds,''
J. Geom. Phys. \textbf{58} (2008) 1662-1692,
{\tt arXiv:0704.1761 [hep-th]}.

\bibitem{Astashkevich:1993ks}
A.~Astashkevich and V.~Sadov,
``Quantum cohomology of partial flag manifolds f(n1 ... n(k)),''
Commun. Math. Phys. \textbf{170} (1995) 503-528,
{\tt arXiv:hep-th/9401103 [hep-th]}.

\bibitem{cf1} I.~Ciocan-Fontanine, ``On quantum cohomology rings of
partial flag varieties,''
Duke Math. J. {\bf 98} (1999) 485-524,
{\tt arXiv:math/9710213 [math.AG]}.

\bibitem{gk} W.~Gu, E.~Kalashnikov,
``A rim-hook rule for quiver flag varieties,''
{\tt arXiv:2009.02810 [math.AG]}.

\bibitem{mihalcea:giambelli} Leonardo C. Mihalcea, ``Giambelli formulae for the equivariant quantum cohomology of
the {G}rassmannian,'' 
Trans. Amer. Math. Soc., {\bf 360} (2008), no. 5, p. 2285-2301,
{\tt arXiv:math/0506335 [math.AG]}.
              
\bibitem{fulton.pandharipande} William Fulton, Rahul Pandharipande,
 ``Notes on stable maps and quantum cohomology", 
pp. 45-96 in {\it Algebraic Geometry --- Santa Cruz 1995},
{Proc. Sympos. Pure Math.} {\bf 62}, 
{Amer. Math. Soc.}, {Providence, RI}, 1997.

\bibitem{fulton:IT} W. Fulton, {\it Intesection theory}, Springer-Verlag, 1984.
	
\bibitem{weihong} W. Xu, ``Quantum K-theory of incidence varieties,''
{\tt arXiv:2112.13036 [math.AG]}.


\bibitem{Iritani:2013qka}
H.~Iritani, T.~Milanov, and V.~Tonita,
``Reconstruction and convergence in quantum {$K$}-theory via
difference equations,''
Int. Math. Res. Not. IMRN (2015) no. 11, 2887-2937,
{\tt arXiv:1309.3750 [math.AG]}.

\bibitem{mihalcea:eqqhom}
L.~Mihalcea,
``On equivariant quantum cohomology of homogeneous spaces:
{C}hevalley formulae and algorithms,''
Duke Math. J. {\bf 140} (2007) 321-350.

\bibitem{lenart.postnikov:affine}
C.~Lenart and A.~Postnikov,
``Affine {W}eyl groups in {$K$}-theory and representation
theory,''
Int. Math. Res. Not. IMRN (2007) no. 12, 1073-7928,
{\tt arXiv:math/0309207 [math.RT]}.






\end{thebibliography}
\end{document}